# Thickness and twist angle dependent interlayer excitons in metal monochalcogenide heterostructures

Wenkai Zheng[§,⊥], Li Xiang[§,⊥], Felipe de Quesada[ℷ,£], Mathias Augustin[ι,l], Zhengguang Lu[§,⊥,†],

Matthew Wilson[ƒ], Aditya Sood[ℷ,£], Fengcheng Wu[∨], Dmitry Shcherbakov[¶], Shahriar Memaran[§,⊥],

Ryan E. Baumbach[§,⊥], Gregory T. McCandless[ϟ], Julia Y. Chan[ϟ], Song Liu[⊤], James Edgar[⊤], Chun

Ning Lau[¶], Chun Hung Lui[ƒ], Elton Santos[ι,l,σ], Aaron Lindenberg[ℷ,£], Dmitry Smirnov[§], and Luis

Balicas[§,⊥]*

E-mail: balicas@magnet.fsu.edu

[§]National High Magnetic Field Laboratory, Tallahassee, FL 32310, USA.

[⊥]Department of Physics, Florida State University, Tallahassee, FL 32306, USA

[ℷ]Stanford Institute for Materials and Energy Sciences, SLAC National Accelerator Laboratory, Menlo Park, CA 94025, USA

[£]Department of Materials Science and Engineering, Stanford University, Stanford, CA 94305, USA

[ι]Institute for Condensed Matter Physics and Complex Systems, School of Physics and Astronomy, The University of Edinburgh, Edinburgh EH9 3FD, UK.

[ƒ]Department of Physics and Astronomy, University of California, Riverside, CA 92521, USA.

[∨]School of Physics and Technology, Wuhan University, Wuhan, 430072 China.

[¶]Department of Physics, The Ohio State University, Columbus, OH 43210, USA.

[ϟ]Department of Chemistry and Biochemistry, Chemistry and Biochemistry Baylor University, Waco, TX 76798, USA.

[⊤]Tim Taylor Department of Chemical Engineering, Kansas State University, Manhattan, KS 66506, USA.

[l]Higgs Centre for Theoretical Physics, The University of Edinburgh, Edinburgh EH9 3FD, UK.

[σ]Donostia International Physics Centre, 20018, Donostia-San Sebastian, Spain.

[†]Present address: Department of Physics, Massachusetts Institute of Technology, Cambridge, MA 02139, USA.

ABSTRACT: Interlayer excitons, or bound electron-hole pairs whose constituent quasiparticles are located in distinct stacked semiconducting layers, are being intensively studied in heterobilayers of two dimensional semiconductors. They owe their existence to an intrinsic type-II band alignment between both layers that convert these into p-n junctions. Here, we unveil a pronounced interlayer exciton (IX) in heterobilayers of metal monochalcogenides, namely γ-InSe on ε-GaSe, whose pronounced emission is adjustable just by varying their thicknesses given their number of layers dependent direct bandgaps. Time-dependent photoluminescense spectroscopy unveils considerably longer interlayer exciton lifetimes with respect to intralayer ones, thus confirming their nature. The linear Stark effect yields a bound electron-hole pair whose separation $d$ is just $(3.6 \pm 0.1)$ Å with $d$ being very close to $d_{Se} = 3.4$ Å which is the calculated interfacial Se separation. The envelope of IX is twist angle dependent and describable by superimposed emissions that are nearly equally spaced in energy, as if quantized due to localization induced by the small moiré periodicity. These heterostacks are characterized by extremely flat interfacial valence bands making them prime candidates for the observation of magnetism or other correlated electronic phases upon carrier doping.

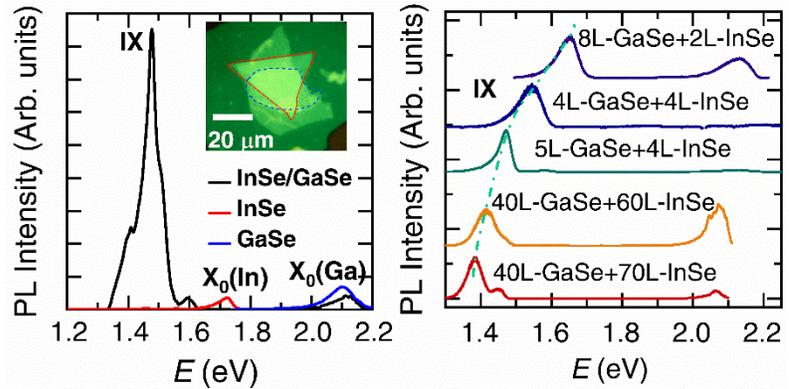

Long-lived excitons, or bound electron-hole pairs created by promoting electrons from the valence to the conduction band, have been proposed as possible building blocks for coherent quantum many-body condensates[1, 2] or even as quantum information carriers[3]. In conventional semiconductors, the exciton lifetime can be increased by building double–quantum wells, where spatially separated electrons and holes become bound to form so-called interlayer excitons (IXs)[4]. Strongly bound IXs can also be formed by stacking, for instance, two single unit cells of transition metal dichalcogenides (TMDs) forming a van der Waals (vdW) heterostructure. Such heterostructures, $e.g.$, $MoSe_2$ on $WSe_2$, or $MoS_2$ on $WS_2$, etc., display ultrafast charge transfer[5], IXs with binding energies as large as ~150 meV[6], and an ability to diffuse over large distances[7]. Furthermore, their tight binding and small Bohr radius can potentially lead to quantum degeneracy between such composite bosons leading to exciton condensation at substantially elevated temperatures when compared, for instance, with those required for conventional Bose-Einstein condensation in cold atoms[2].

In TMD monolayers the spatial separation between bound charge carriers in distinct layers yields distinct properties to IXs when compared to those of the intralayer excitons ($X_0$), due to the reduced overlap between electron and hole wavefunctions. In quantum wells the lifetime of their indirect IX greatly exceeds that of the direct excitons[8], allowing them to thermalize with the lattice prior to recombination. Furthermore, the electron–hole separation results in a permanent electric dipole moment, which creates repulsive dipole–dipole interactions[9]. In stacked TMD monolayers the IXs possess properties that result from the vdW coupling between layers, enhanced Coulomb interactions at the 2D limit, in combination with the intrinsic valley properties of the constituent layers leading, $e.g.$, to locked electron spin and layer indices [10]. In effect, a lattice mismatch between hetero-bilayers, or a small twist angle between homo-bilayers, leads to a long range periodicity or moiré pattern, that affects their overal optical response and therefore the properties of the excitons in 2D vdW semiconducting layers [11-17].

Scanning tunneling microscopy (STM) performed on $MoS_2$/$WSe_2$ heterobilayers[18], reveals a periodically modulated local band gap within the moiré unit cell. This forms a two-dimensional electronic superlattice whose electronic wavefunctions are spatially localized within the potential minima of the individual moiré unit cells. Moiré quantum systems display electronic, optical and topological properties that are tunable with an exquisite level of control[11, 13-17, 19-23]. However, to date the moiré physics has been constrained by two factors i) the dimensionality, defined by the stacking of monolayers[19], and ii) the twist angle $\phi$, which unveils superconductivity[21], orbital magnetism[22], correlated insulator states[17, 20] only at precise values of $\phi$. The moiré potential in $WSe_2$/$WS_2$ heterostructures is much stronger than the exciton kinetic energy, and generates multiple flat excitonic minibands that strongly alter the intralayer exciton emissions of $WSe_2$[13]. TMD monolayers have a direct band gap located at the degenerate $K$, $K$' points at the corners of their unfolded Brillouin zone. However, efficient light emission from IX recombination requires bound electron-holes to emerge from almost the same momentum in reciprocal space. In twisted TMD heterobilayers, a pronounced IX emission is observed only for twist angles approaching either $0^o$ or $60^o$ due to a momentum mismatch between the minimum in energy of the conduction band and the valence band maximum at larger twist angles. This mismatch suppresses emissions at intermediate angles[24] where umklapp recombination processes contribute to a weak emission[15].

In this work, we introduce another platform to explore the physics of the interlayer excitons and possibly the role of the moiré potential on the optoelectronic response of layered semiconductors: stacked semiconducting metal-monochalcogenides (MMCs) or γ-InSe/ε-GaSe heterostructures fabricated under inert atmosphere (see, Methods). Similar to TMDs, the band gaps of both ε-GaSe and γ-InSe progressively increase as the number of layers is reduced[25] albeit evolving from direct to indirect. Bulk γ-InSe displays a direct bandgap of 1.25 eV, but transitions to a 2.9 eV indirect one in the single layer limit[26] while the 2 eV band gap of bulk ε-GaSe increases up to ~2.4 eV in bilayers[25] (Fig. 1). This behavior implies that the relative type-II band alignment[27] between γ-InSe and ε-GaSe can be modified simply by stacking layers having distinct thicknesses. This provides an additional level of tunability for the energy of the interlayer exciton in γ-InSe/ε-GaSe heterostructures and contrasts with the TMDs that display a direct band gap only at the monolayer limit[28]. In addition, the direct band gap of both compounds, excepting perhaps for their bilayers and monolayers, is at the at the Γ-point of their Brillouin zone[25]. This absence of momentum mismatch between electrons and holes in adjacent layers, regardless of their twist angle, would represent an advantage with respect to TMDs for studying interlayer excitonic emissions as well as the possible effect of the moiré potential on them. In MMCs, band structure calculations indicate that optically induced interlayer electronic transitions would always be direct in k-space (except for their monolayers)[25, 29]. The fact that InSe/GaSe heterostructures form type-II band alignments despite variations in the thicknesses of the constituent layers, was already discussed in several reports[25, 27, 30, 31].

Here, we show through photoluminescense measurements that InSe/GaSe heterostructures encapsulated among h-BN layers display a very pronounced and multicomponent interlayer exciton emission whose energy can be tuned by varying the thickness of the constituent layers. The lifetime of the interlayer exciton is considerably longer than those of the constituent layers, as observed in TMDs. Under an electric field, the associated Stark effect is dominated by the linear term yielding a dipole separation of just $(3.6 \pm 0.1)$ Å. When the IX emission is decomposed into its components, the higher energy superimposed emissions also display the second-order Stark effect, with both observations being akin to those from hydrogen atoms, suggesting relatively isotropic wave-functions for IXs. Finally, we show that the multicomponent IX emissions display similarities with moiré exciton behavior in TMDs, namely twist angle dependent energy quantization that is reproducible in every location across the interface. This would suggest small radii for the interlayer excitons in metal monochalcogenides when compared to those of TMDs[32].

RESULTS AND DISCUSSION

Thickness dependent photoluminescence in γ-InSe/ε-GaSe heterostructures

The heterostructures were fabricated from exfoliated high quality bulk single-crystals of γ-InSe and ε-GaSe. Crystal synthesis protocols are described in Methods. Our γ-InSe crystals yield quantum Hall effect and mobilities exceeding 4000 cm²/Vs [33]. Single-crystal x-ray diffraction (Fig. S1) reveals no stacking mosaicity in the ε-GaSe. A simple schematics of our heterostructures (Fig. 1a) illustrates that the individual layers are not grounded during the measurements and therefore cannot be doped upon gating. Few layered graphite was transfered on top of the stack to act as the top electrical gate. InSe and GaSe share a planar hexagonal structure, but with distinct lattice constants of $a_0 = 4.08(2)$ Å and $a = 3.7557(3)$ Å, respectively. This marked difference is sufficient to generate a moiré modulation even when both layers are perfectly aligned. For twisted layers, the moiré periodicity λ as a function of twist-angle $\phi$ would range from a mere λ ~ 5 nm ($\phi$ = 0º) to just 0.7 nm ($\phi$ = 30º), see Ref. [34]. If the Bohr radius of IX was smaller than λ, IX would sense the moiré potential and become confined by it. In fact, the separation and confinement of the bound electron-hole pairs in spatially separated layers is expected to make IXs particularly susceptible to the moiré potential at the interface[13].

The type-II band alignment between both compounds, leading to a near direct band gap at the $\Gamma$-point, is supported by band structure calculations for monolayers (Fig. 1b) as well as few layered crystals having a similar number of layers L (Fig. S2). To understand the interfacial electronic structure, we performed van der Waals (vdW) corrected ab-initio simulations finding a large interfacial charge density involving the Se atoms of both materials; they are separated by just 3.4 Å (Fig. 1c). This proximity between interfacial Se-Se atoms induces a strong interlayer coupling of near covalent character that maximizes the role of the moiré potential, or leads deeper moiré potential wells relative to conventional vdW coupled layers. The calculated electronic charge densities (Fig. S3), and associated potential difference (Fig. S4) leads to a pronounced interfacial electric field (Fig. S5). Our calculations are supported by previous scanning photoelectron microscopy measurements in $\gamma$-InSe/$\varepsilon$-GaSe heterostructures that reveal a charge dipole at the interface, with the conduction band offsets suggesting that charge carriers are able to overcome the built-in potential to meet at the interface and form an interlayer exciton[30].

To evaluate the optical response of these interfaces we collected micro-PL spectra from several $\gamma$-InSe/$\varepsilon$-GaSe heterostructures whose constituent layers vary widely in their thicknesses. The samples were kept at $T$ = 5 K and photoexcited with a continuous-wave $\upsilon$ = 514-nm laser ($\upsilon$ is the wavelength) focused through an objective (of numerical aperture NA = 0.65) to a spot size of ~2 μm. For instance, we collected photoluminescence (PL) spectra from a 4L-$\gamma$-InSe flake on 5L-$\varepsilon$-GaSe (Fig. 1d) revealing that the intensity of the IX emission from the interfacial area is considerably more pronounced than those of the intralayer $X_0$(In) and $X_0$(Ga) emissions collected outside of the interface, despite their direct band gap. In addition, the intensity of $X_0$(In) collected in the interfacial area becomes nearly completely suppressed. A similar quenching occurs in TMDs and is attributed to ultrafast interlayer charge transfer[5, 13, 15]. This quenching is observed in nearly all measured heterostructures, with thicknesses ranging from $n$(In) $\cong$ 2L to $n$(In) $\cong$ 72L for γ-InSe, and from $n$(Ga) ranging from $\cong$4L to $\cong$ 38L  (Fig. 1e).

The IX emission observed at 1.38 eV in thick constituent layers, increases up to 1.65 eV (indicated by the dotted line) as their thicknesses decrease or as their band gap increases. Notice the relative agreement between these values and the calculated energy separation between valence and conduction bands  (Fig. 1b and Fig. S2). As previously mentioned, Figs. S6, S7, S8 display a comparison between emissions from the interfacial area and those from the intralayer excitons associated exclusively with the γ-InSe and ε-GaSe areas located beyond the interface.   The important point is that this pronounced IX peak does not coincide with either $X_0$(In) or $X_0$(Ga), albeit for thick layers it partially overlaps with $X_0$(In). Therefore, it ought to correspond to emissions from interlayer excitons (here, the term "interlayer" means distributed across both materials). From the overall data collected by us, we observe that to first approximation the energy of the interlayer exciton $E$(IX) follows a power law dependence on both $n$(In) and $n$(Ga), i.e., $E$(IX) $\propto [n(Ga,In)]^{\alpha}$, with exponents $\alpha$(In) = -(35 $\pm$ 9) and $\alpha$(Ga) = -(30 $\pm$ 4). This leads to the expression $E$(IX) = $a[n(Ga)]^{\alpha(Ga)} + b[n(In)]^{\alpha(In)}$, with coefficients $a \cong$ 2.52 eV and $b$ = -0.86 eV, providing a rough estimate for the value of $E$(IX) as a function of either layer thickness.

In Fig. S7 we have replotted the PL spectra displayed in Fig. 1e, fitting them to a set three Gaussian functions that allow us to decompose the IX emissions. Fig. S7 also allows a closer inspection of the precise position in energy of the IX emissions from each heterostructure. As seen, most IX emissions can be well fitted to just three Gaussians whose separation in energy is heterostructure dependent. For several heterostructures we also detect the intralayer exciton $X_0(In)$ in the vicinity of IX. The important observations are i) that the separation in energy $\Delta E$ between all three IX emission is nearly constant, as is the case for the few heterostructures discussed below, ii) but the separation in energy between the highest IX emission and the one from $X_0(In)$ is considerably larger and tends to show a much lower amplitude, allowing us to differentiate them. As discussed below and in the SI, the emissions yielding IX do not match the $X_0(In)$ ones measured at low temperatures for bulk $\gamma$-InSe implying that they have a distinct origin.

As seen in Figs. S6, S7 and S8, both the $X_0(In)$ and $X_0(Ga)$ emissions become strongly quenched in the junction area, but this quenching is particularly pronounced for samples characterized by $\phi < 20^\circ$. This implies that $\phi$ affects the charge transfer across the interface with this transfer becoming particularly fast or efficient at smaller values of $\phi$. Throughout this text, the values of $\phi$ were determined through second harmonic generation (see, Figs. S9 through S14).

Evaluation of exciton lifetimes through time-dependent PL

In TMDs, the IX emissions display considerably longer lifetimes relative to the intralayer ones[6, 14, 35, 36], with their lifetimes being modulated by factors like the incident illumination power[37] or the twist angle between heterobilayers[38], i.e., as the direct optical transition (for $\phi = 0^\circ$) at the $K$-point becomes indirect upon increasing $\phi$ and leading to excitons with a center of mass momentum[38]. The measured IX lifetimes $\tau$ varies widely[6, 14, 35, 36-39], from values ranging from $\tau \cong 10^{-1}$ ns up to $\tau \cong 10^2$ ns. The reason for this variability in the values of $\tau$ remains unclear, although the superposition of an additional indirect optical transition when $\phi > 0^\circ$ or the scattering between bright and dark excitons may explain heterostructure dependent values for $\tau$ [38, 40]. These same factors might also explain the characteristic two exponential decaying times of the photoluminescent response [36, 38].

To further characterize the nature of the IX emission, we have also conducted time dependent photoluminescense measurements (Fig. 2) in a 4L-$\gamma$-InSe / 4L-$\varepsilon$-GaSe heterostructure that is characterized by $\phi$ = (9.4 ± 0.6)° (Fig. S10). This heterostructure was chosen because it displays clear IX, $X_0$(In), and $X_0$(Ga) emissions at 1.52 eV and ~1.7 eV ~ 2.12 eV, respectively (Fig. 2a). Each emission can be well adjusted to two exponential decays once the instrumental response function is subtracted from the experimentally determined time-dependent photoluminescent signal (Fig. 2b). These fits yield values $\tau_1$ for the shortest decay time of 0.0513 ns, 0.0632 ns, and 0.2808 ns for $X_0$(Ga), $X_0$(In), and IX, respectively. And $\tau_2$, or longer decay times of 2.2030 ns, 0.9739 ns, and 2.7546 ns for $X_0$(Ga), $X_0$(In), and IX, respectively. These values are comparable or shorter than the exciton lifetimes reported for TMDs[6, 14, 35, 36-39]. Most importantly, both lifetimes extracted for IX are considerably longer than those associated with the intralayer excitons, as previously reported for TMDs, thus supporting our identification of IX as an interlayer exciton whose emission energy is layer thickness dependent. We also evaluated the $\tau_1$ and $\tau_2$ lifetimes associated to the IX emissions extracted from thicker heterostructures characterized by $\phi \cong$ 30° and 40°, obtaining respectively, $\tau_1$= 0.1522 ns and 0.1275 ns, or $\tau_2$ = 1.9396 ns and 2.2673 ns. Hence, larger twist angles would seem to affect $\tau_1$ considerably in contrast to $\tau_2$. A considerable increase in exciton relaxation times as a function of twist angle has already been reported for TMDs[38]. However, to describe the exciton decay times in MMCs, one would have to understand their ground and excited excitonic states, how these couple to acoustic and optical phonons, and how this coupling might be affected by the twist angle. Prior to this, a systematic study in multiple samples will have to be performed to clarify if the differences in $\tau_{\mathrm{I}}$ are indeed twist angle dependent or sample dependent.

Stark effect

The envelope of IX at any given location should be controllable through the application of an electric field given that interlayer excitons possess an out-of-plane electric dipole moment that is manipulable *via* an out-of-plane electric field[37], or the Stark effect. For example, the IX emissions from a heterostructure composed of 5L-ε-GaSe on 4L-γ-InSe, fitted to a mimimum set of three Lorentz functions, are blueshifted by approximately 25 meV as the top gate voltage $V_{\mathrm{TG}}$ (Figs. 3a to 3c) is swept from -28 to 28 V. The stark Hamiltonian, $H_{\mathrm{Stark}} = \mu E$, descripts the coupling between the electric field $E$ and the exciton electric dipole moment $\mu = ed$ oriented across the heterostructure ($d$ is the electron-hole separation). The linear Stark-effect is expected in our case, given the lack of inversion symmetry at the interface and the orientation of the dipole moments. Applying an $E$ antiparallel to $\mu$ affects the overlap between electron and hole wavefunctions and therefore the PL emission (as seen in Fig. 3b). Similar behavior was reported for twisted TMD heterobilayers[41], although it was found to be non-linear and far more pronounced in TMD homobilayers[42]. A linear fit between the electric field applied across the heterostructure $E_{\mathrm{h}}$ and the energy of the lowest energy emission (centred at ~1.42 eV at $V_{\mathrm{TG}} = 0$ V) yields $d = (3.6 \pm 0.1)$ Å. This value is close to our DFT calculated separation between adjacent interfacial Se atoms, or 3.4 Å (Fig. 1c), thus experimentally confirming our prediction of charge accumulation in the vicinity of the Se atoms. The higher energy IX emissions display a mild non-linear dependence on $E_{\mathrm{h}}$ (Fig. 3c) that is well captured by a *negative $E^2$* term associated to the second order Stark-effect. Note, that a linear response on electric field for the lower energy emissions, with higher energy ones revealing the superposition of a quadratic term, is precisely the behavior of the hydrogen atom where the electrical polarizability of its atomic orbitals was empirically determined to be proportional to $n$ the principal quantum number[43].

The linear dependence on $E_h$ for IX contrasts markedly with the one displayed by the intralayer exciton emission of 5L-$\gamma$-InSe (Fig. 3d) which displays a mirror symmetric, purely quadratic response with respect to $E_h = 0$ V. This implies that the $X_0(In)$ does not possess an intrinsic dipole moment, in contrast to IX, or that its dipole moment is induced by $E_h$ due to a symmetric and nearly spherical distribution of charges in the absence of the electric field. This contrasting behavior between $X_0(In)$ and IX further supports the notion of IX being indeed an interlayer exciton emission from the interface between both compounds.

Evidence for energy quantization

Having provided enough experimental evidence to support the presence of a pronounced interlayer exciton emission IX in $\gamma$-InSe/$\varepsilon$-GaSe heterostructures, we proceed to discuss its structure which, as shown below and in sharp contrast to the emissions discussed in Fig. S7, is composed of several superimposed emissions which are nearly equally spaced in energy, as if quantized. Energy quantization would result from excitons localized at the moiré potential minima of twisted bilayers[11, 14, 40]. Nevertheless, despite the number of reports claiming to observe so-called moiré excitons, the community is not yet in position to discard other possible interpretations, such as phonon-mediated states, defect-bound states, or spin triplet states [37–40][44]. Unambiguous evidence for moiré trapped excitons would require the ability to image the optical response of individual mini moiré Brillouin zones, and confirm its reproducibility throughout the entire interfacial area of the heterostructure. Ours, as well as the experimental set-ups used by other groups, focus the impinging light excitation into spots having a typical diameter of ~ 1 μm. This spot area is considerably larger than the typical moiré periodicity λ, which ranges between one and tens of nanometers, implying that the incoming light excites thousands of moiré unit cells. Nevertheless, we show below that the structure of IX, collected from different points at the interface, is quite reproducible

A sketch of a 1L-γ-InSe stacked onto 1L-ε-GaSe forming a small $\phi$ between their main crystallographic axes  (Fig. 4a) illustrates the moiré periodicity λ resulting from modulations in the local stacking. According to Ref.[34], for our heterostructures λ would range from ~46 Å (for $\phi$ = 0º) to ~7 Å (for $\phi$ = 30º) (Fig. 4b). To support our claim about the interlayer nature of the IX exciton emission centred at ~1.5 eV for a 5L-ε-GaSe/4L-γ-InSe heterostructure, we also measured its amplitude as a function of the laser excitation energy, i.e., from ~1.6 to 2.6 eV (Fig. 4c). The amplitude of the IX emission increases as the excitation energy surpasses 1.7 eV (corresponding to the $X_0$(In) for 4L-γ-InSe) reaching a plateau beyond ~ 2.1 eV (the $X_0$(Ga) emission for 5L-ε-GaSe). This implies that intralayer excitons from both layers contribute to IX as they decay. However, we did not observe clear PL peaks at the absorption energies corresponding to each intralayer emission. The reason for this is not clear, although it points to very fast intralayer exciton decays *via* interlayer recombination processes.

As mentioned previously, the IX emission from twisted γ-InSe/ε-GaSe heterostructures can be decomposed into several peaks. This implies the superposition of multiple emissions as claimed for excitons trapped by the moiré potential in twisted TMDs[11, 13-17,16]. To illustrate this point, we fit the IX emissions (Figs. 4d, 4e and 4f), from three distinct heterostructures characterized by different constituent layer thicknesses and twist angles, i.e., $\phi = 0.3°$, 23.4°, and 47.9° as determined through second harmonic generation measurements (Figs. S9 to S14), to a minimum set of three Gaussians that satisfactorily describe these IX emissions. The reason for choosing 3 Gaussian functions is illustrated by the fits in Fig. S7: three Gaussians are needed for a proper fit of the IX emissions while also capturing in certain heterostructures the very weak $X_0(In)$ emission in the vicinity of IX. In contrast to the PL spectra shown in Figs. S7d and S7e whose $X_0(In)$ emissions are not only weak but also well separated from IX, for the three heterostructures whose deconvoluted IX emissions are shown in Figs. 4d to 4f, $\Delta E$ remains nearly constant as if resulting from energy quantization. $\Delta E$ increases from an average value of ~15 meV to ~52 meV as $\phi$ increases from 0.3° to 47.9° indicating that their envelopes broaden considerably upon increasing $\phi$, implying that they are particularly sensitive with respect to the twist angle. These observations would be compatible with energy quantization due to exciton localization induced by a moiré potential whose periodicity $\lambda$ is defined $\phi$. Notice that the values of $\Delta E$ extracted here are comparable to $\Delta E \sim 20$ meV typical for the IX emissions from TMD heterobilayers that are indeed characterized by a moiré periodicity[14]. In Fig. 4, the reason for including results from heterostructures assembled from layers having quite disparate thicknessess, e.g., Figs. 4e and 4f, is to clearly illustrate two points: i) the position in energy of the IX emission is clearly layer thickness dependent and ii) that the overall width of the IX emission is clearly twist angle dependent.

Importantly, our PL measurements performed at different locations throughout the interface reveal that the envelope of the IX emission is position dependent, but always well described by the same set of Gaussians with nearly constant $\Delta E$ (Fig. 4f). In a moiré interlayer exciton scenario, position dependent emission amplitudes would be compatible with mild local strain[45], or modulations in the inter-layer coupling associated to lattice relaxation that modulates their coupling to the moiré potential. Disorder, adsorbates, and polymer residues, would be expected to produce random, position dependent emissions in contrast to what is exposed here, and to what was claimed for moiré excitons in TMDs[15]. One must bear in mind that the displacement of IX to higher energies (Fig. 1e) as the thicknesses of the constituent layers decrease, or as the $\gamma$-InSe and $\varepsilon$-GaSe direct gaps increase[25], can only be reconciled with the IX emissions being intrinsic to the heterostacks instead of (random) emissions from defects.

We also argue that these IX emissions cannot be associated with bulk emissions resulting from, for instance, phonon replicas or other bulk processes[46]. To support this argument, we evaluated the PL spectra from bulk $\gamma$-InSe at $T = 300$ K as well as $T = 5$ K (Fig. S15), from encapsulated 5L-$\gamma$-InSe (Fig. S16), and bulk $\varepsilon$-GaSe at $T = 5$ K (Fig. S17). In all cases, the amplitude of the individual intralayer emissions (fitted to Lorentzian functions) increase as their energy increase, in sharp contrast to what is seen in Figs. 4d through 4f for IX. In addition, the emissions from $\varepsilon$-GaSe are observed to be *not* equidistant in energy, as in the case for those from IX, being observed at much higher energies than the IX emissions. Furthermore, at $T = 5$ K the emissions from $\gamma$-InSe become considerably narrower than those from IX seen in Figs. 4d through 4f. In addition, and as seen in Fig. S16, the individual emissions from bulk and thin exfoliated $\gamma$-InSe are considerably sharper, i.e., ~ 14 meV at FWHM for the bulk, and ~ 5 meV at FWHM for the exfoliate crystal, with respect to those of IX shown in Fig. 4, which range from 20 meV and 50 meV at FWHM. All these observations, indicate that the splliting of the IX emissions have a distinct origin with respect to the emissions observed in the original compounds.

Notice that by moving the laser spot to the edge of the 5L-γ-InSe encapsulated crystal we observe fairly narrow random emissions (whose full width at half maximum is < 1 meV) that are akin to those seen in Ref.[15] which, in contrast, we associate to edge defects or broken bonds resulting from the exfoliation process. Nevertheless, these IX emissions emerge only below $T$ < 80 K (Fig. S18) confirming that they are not intrinsic to the original bulk compounds. However, IX blue shifts and saturates as the illumination power increases (Fig. S19) and this is precisely what is expected for interlayer excitons dominated by Coulomb like dipole-dipole interactions[37].

To first approximation, the energy quantization of the IX emissions can be understood in terms of excitons trapped by a parabolic potential (describing, e.g., the confinement within a moiré potential minima). Excitons can be treated as wave-packets that move through a "pinning" potential with the motion of their center of mass described by[14]:

$$V = V_0 + \hbar^2 k^2/2M + V_m(x,y) \qquad (2)$$

Here, $V_0$ is a constant given by the energy gap associated with the type-II band alignment, $\hbar^2 k^2/2M$ is the kinetic energy of the exciton center of mass, and $V_m(x,y)$ describes the potential minima (e.g., produced by the moiré periodicity). In the strong coupling regime, $V_m(x,y) = \frac{1}{2}m\omega_0^2(x^2 + y^2)$, where $\omega_0^2$ is the eigenfrequency within the harmonic trap and $m$ is the exciton effective mass. Such potential would lead to twist angle dependent eigenenergies that would indeed be quantized, as seen experimentally[14]:

$$E_{n_x,n_y}(\phi) = E_0 + h\omega_0(\phi)\big[n_x + n_y\big] \qquad (3)$$

Although the evidence provided here is consistent with localized excitons displaying quantized energy levels, we are not yet in position to solidly claim that this arises from the moiré periodicity. This would require a precise determination of the IX planar radius *via*, for example, measurements of the diamagnetic shift under very high magnetic fields[47], to compare it with the moiré periodicity $\lambda(\phi)$ (Fig. 4b).

Notice that a measured dipole separation $d \cong 0.36$ nm is considerably smaller (by nearly one order of magnitude) than the typical IX radii ($\sim 1.6$ nm) extracted from TMDs encapsulated among *h*-BN layers[32]. This dimension is also smaller than the calculated moiré periodicity as a function of $\phi$ (Fig. 4b), indicating that a nearly isotropic *s*-wave exciton could become pinned by the potential minima associated to the rather small moiré periodicities intrinsic to the γ-InSe/ε-GaSe heterostructures. Nevertheless, more work is needed to verify this tantalizing possibility by focusing on a comparison between the planar dimensions of IX relative to the moiré periodicity λ of a given heterostructure. We finalize by mentioning that the intralayer excitons have also been reported to be particularly susceptible to the moiré potential[13]. In fact, there are multiple superimposed emissions also for $X_0$(Ga) that are fairly reproducible throughout the interface between both compounds (Figs. S19 and S20).

CONCLUSIONS

In summary, we report a pronounced interlayer exciton emission in metal monochalcogenide heterostructures at low temperatures, whose energy is layer thickness dependent and whose width (in energy) is twist angle dependent. The overall phenomenology unveiled here is similar to previous reports on excitons subjected to the moiré potential intrinsic to twisted bilayers of transition metal dichalcogenides. However, and in contrast to TMDs, this very pronounced interlayer emission occurs over the entire range of twist angles and a broad variation in layer thicknesses. We argue that its existence is attributable to the direct band gap at Brillouin zone center that is intrinsic to multilayered metal monochalcogenides.

At this point, it is quite important to clarify the origin of its quite reproducible multicomponent emissions. These cannot be attributed to phonon replicas given that replicas would not lead to nearly equidistant (in energy) emissions, or defects and adsorbates at the interface that would lead to random emissions as a function of interfacial spatial coordinates, in clear contrast to our observations. Should these emissions result from exciton localization due to the underlying moiré potential it would imply the existence of interlayer excitons with small radii in metal monochalcogenide heterostructures.

In effect, as discussed in ref. 18, an uniform and ordered array of nanodots (*e.g.*, excitons trapped by a moiré potential) could be utilized for semiconductor lasers or as single-photon emitters. Artificial lattices of interacting quasiparticles also offer a potential for applications in quantum information processing, through the the creation of coherent exciton condensates, with γ-InSe/ε-GaSe heterostructures offering a high level of tunability of their optical response with an intrinsic electric field at their interface that would find natural applications in, for instance, photovoltaics. Upon charge carrier doping, it might be possible to induce a complex, twist angle dependent phase-diagram containing correlated electronic phases[48]. Furthermore, the optical tunability intrinsic to metal monochalcogenides can be further enhanced by creating stacks with their S or Te analogs, or even transition metal dichalcogenides[49].

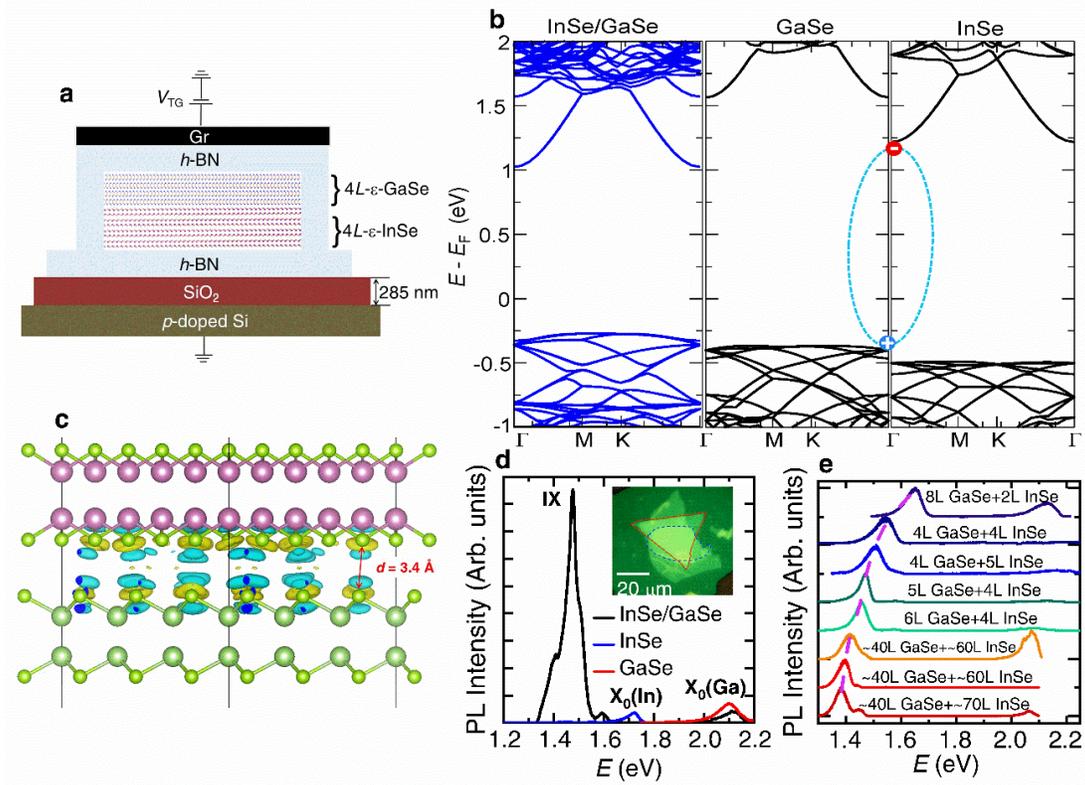

**Figure 1. Photoluminescence spectra from γ-InSe/ε-GaSe heterostructures displaying inter-layer exciton emission.** (a) Sketch of a typical heterostructure based on $n$(Ga)$L$-ε-GaSe on $n$(In)$L$-γ-InSe encapsulated among $h$-BN layers (clear blue) with the entire stack transferred onto SiO₂ (granate). A top thin graphite (Gr) layer (thickness ∼ 100 nm) was transferred onto the top of the stack to act as the top electrical gate with $h$-BN acting as the dielectric layer. Notice that the individual layers are not connected to the electrical ground. Therefore, the application of an electric field *via* either the top or the bottom gates cannot charge dope the constituent layers or the interface. (b) Left to right: band structures of a stack composed of monolayer (1$L$) ε-GaSe and 1-$L$-γ-InSe, 1-ε-GaSe (center), and 1-$L$-γ-InSe (right) within the reduced moiré Brillouin zone where the twist angle is $\phi = 0^\circ$. Here, the position of the conduction and valence bands of 1$L$-ε-GaSe (center) relative to those of 1$L$-γ-InSe (right) illustrates their type-II band alignment. The folded and quite flat valence band of 1$L$-ε-GaSe within the moiré Brillouin zone, allows the formation of an interlayer exciton with the electron promoted to the conduction band minimum of 1$L$-γ-GaSe located ~~also~~ at the Γ-point. This lack of momentum mismatch between maximum and minimum is favorable to the creation of interlayer excitons (blue and red dots encircled by dotted line depict the interlayer exciton at Γ), regardless of the twist angle. (c) Difference in charge density (2 x 10⁵ eV/A³) at the InSe/GaSe interface calculated through vdW-corrected *ab-initio* simulation

methods. The charge density is confined mainly at the interface and involves the Se atoms from both materials, which are separated by 3.4 Å. The small Se-Se distance at the interface (3.4 Å) is close to the lengths of the chemical bonds in In-Se (2.60 Å) and Ga-Se (2.50 Å). This proximity of the Se-Se atoms at the interface induces a covalent character altogether with the vdW interactions. In the plot, big purple (green) atoms depict Ga (In), and small green atoms correspond to Se. In Extended Data Figs. 1, 2 and 3, we provide the calculated electronic charge distribution, the electrical potential and electric field as a function of the interlayer distance $z$, indicating that these heterostructures are form PN-junctions.   (d) Measured PL spectra from a $n$(In) = 4L InSe layer (blue line), a $n$(Ga) = 5L-ε-GaSe (red line), and from their interface (black line). The interfacial exciton IX peaks at a lower energy with respect to the intra-layer excitons $X_0$(Ga) and $X_0$(In) of the individual constituent layers, thus indicating that IX originates from an interlayer transition. Notice its multi-component nature. Inset: optical microscopy image of the $h$-BN encapsulated 4L-ε-GaSe/5L-γ-InSe heterostructure where the GaSe flake is delineated by a solid red line with the InSe one delineated by a blue dashed line. (e) Photoluminescence (PL) spectra for several $n$(Ga)L-ε-GaSe/$n$(In)L-γ-InSe heterostructures, with the number of layers L varying between $n$(Ga) = 4 and 40, and $n$(In) from ~2 to ~70, measured at $T$ = 5 K. In this panel all curves are offset for clarity. Owing to the type-II band alignment between both compounds, the interlayer exciton emission (or IX, indicated by the dotted line) is located between~ 1.39 to and 1.65 eV as the thicknesses of the constituent layers decrease. In contrast, the intra-layer exciton emission for both InSe and GaSe  become strongly quenched in the junction area.

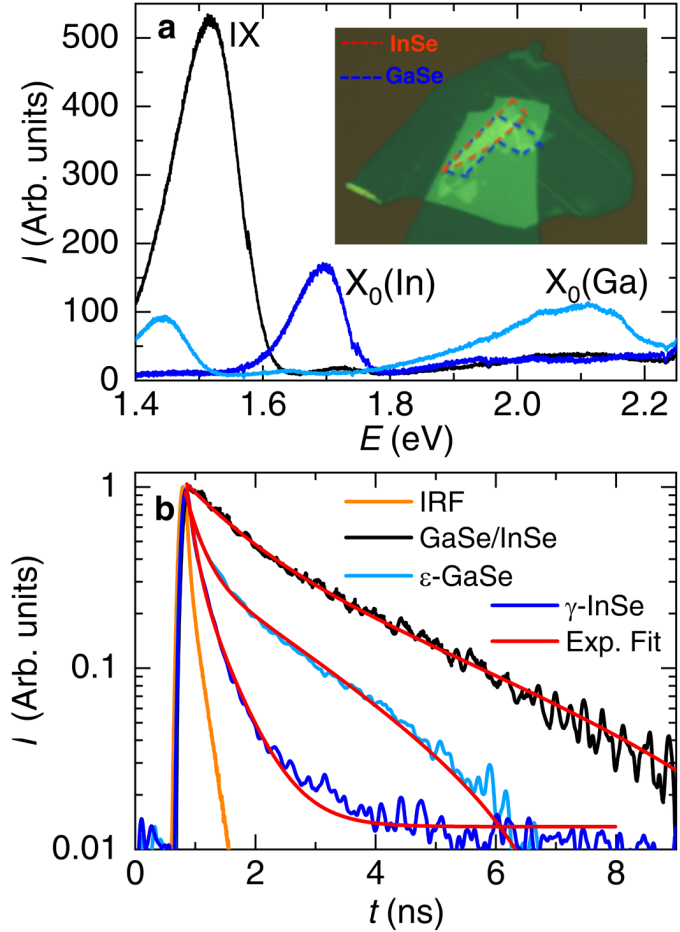

**Figure 2. Interlayer and intralayer exciton lifetimes. (a)** Photominescence spectra from a non-annealed 4-γ-InSe/4L-ε-GaSe heterostructure (see, Fig. S10). Black trace corresponds to the PL spectrum from the junction area, revealing emission peaks at ~1.52 eV and ~1.7 eV as well as a broad peak centred at ~2.12 eV. Blue trace corresponds to emission from a region of the γ-InSe layer that does not overlap the ε-GaSe one, indicating that the peak at ≈1.7 eV corresponds to the intralayer exciton $X_0$(In) of γ-InSe. Clear blue trace depicts the emission from the ε-GaSe layer not in contact with γ-InSe, indicating that the peak at 2.12 eV corresponds to the intralayer $X_0$(Ga) emission. Therefore the tall at peak at 1.52 eV, ought to correspond to the interlayer IX emission. Inset: micrograph of the heterostructure where the boundaries of the γ-InSe and ε-GaSe crystals are indicated by red and blue dashed lines respectively. **(b)** Measurements of the exciton lifetimes through time-dependent photoemission spectroscopy. Here, blue, clear blue, and black traces correspond to $X_0$(In), $X_0$(Ga), and IX respectively. Orange line corresponds to the time dependent Instrument Response Function (IRF) of the set-up used for the measurements. Red lines are fits to two exponential decays, illustrating the existence of two lifetimes. As clearly seen, IX has considerably longer lifetimes with respect to both X$_0$(In) and X$_0$(Ga). The actual values of the lifetimes are obtained by deconvoluting the IRF from the experimental time-dependent photoluminescent signal. The goodness of these fits $R^2$ are 0.99875, 0.99732, and 0.99682 for IX, X$_0$(Ga),and X$_0$(In), respectively.

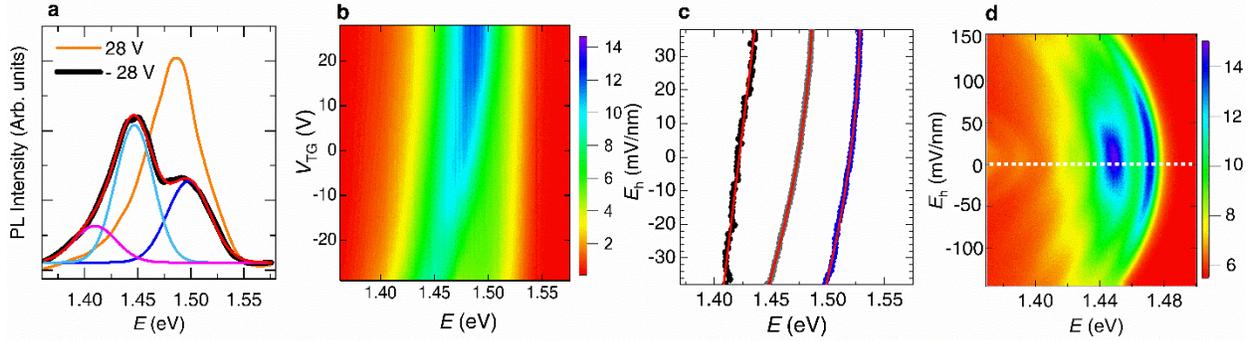

**Figure 3. Role of electric field on both the interlayer and intralayer X$_0$(In) exciton emissions. (a)** Interlayer exciton emission from the 5L-ε-GaSe on 4L-γ-InSe heterostructure ($\phi \cong 23.4^{\circ}$, see Fig. S11) for three values of the top-gate voltage, $V_{TG}$ = +28 V (orange trace), and -28 V (black markers). Magenta, clear blue, and blue lines are Gaussian fits to the interlayer exciton emission under $V_{TG}$ = -28 V. Red line is their supperposition. **(b)** Contour plot depicting the intensity of the interlayer exciton emissions IX illustrating their mild non-linear response (blue shift) as $V_{TG}$ (or electric field) is varied. **(c)** Center of the fitted Gaussians (in eV) as $V_{TG}$ is tuned. The emission at the lowest energy (black markers) can be well-described by a linear fit (red line), while those at higher energies (gray and blue markers) are described by a linear plus a negative quadratic term corresponding to the second-order Stark shift correction (dark magenta lines). Here, the electric field $E_h$ is given by $E_h = V_{TG}/t$ x ε($h$-BN)/ε(InSe-GaSe) where the dielectric constant ε($h$-BN) ≈ ε(SiO$_2$) ≈ 3.7 and ε(InSe-GaSe) is the average value between ε(InSe) = 7.6 and ε(GaSe) = 6.1 or 6.85. From the linear fit we calculate the electric dipolar moment μ$_e$ = $e.d$, yielding an electron-hole separation $d$ = (3.6 ± 0.1) Å, which is close to the calculated value of 3.4 Å for the interlayer Se separation at the interface. This value increases by ∼ 35 % for the emission at the highest energy. **(d)** Intensity of the intralayer exciton emission of InSe, X$_0$(InSe), as a function of $E_h$ for a 5L-γ-InSe crystal encapsulated among $h$-BN layers in a double gated configuration. The intensity of X$_0$(InSe) is symmetric with respect to positive and negative values of $E_h$ displaying a quadratic dependence on $E_h$. Such dependence, which is consistent with the non-linear Stark effect, is observed in nearly isotropic

excitons whose dipole moment is induced by the external electric field. This contrasting behavior between IX and X₀(In) confirms that the former emission cannot be attributed to the original X₀(In) when subjected to the distinct dielectric environment of the heterostructure.

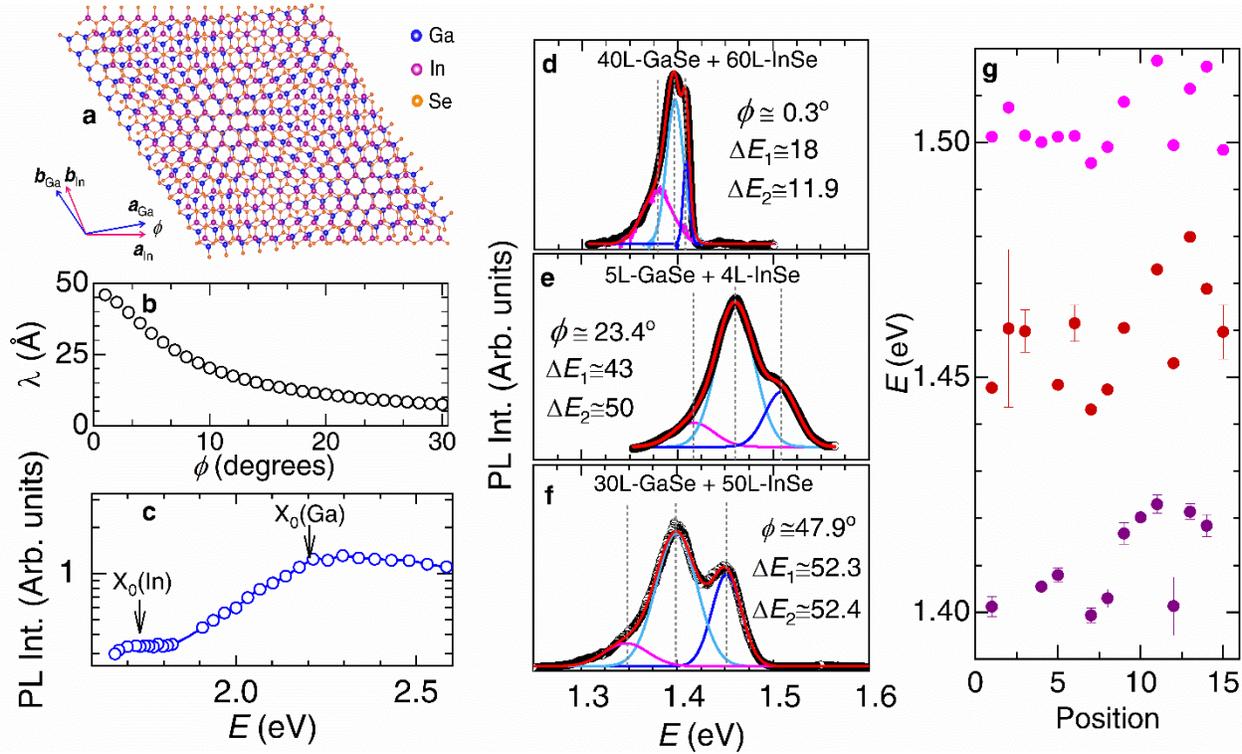

**Figure 4. Twist angle dependent interlayer IX exciton emissions and evidence for energy quantization. (a) Depiction of 1L-γ-InSe superimposed onto 1L-ε-GaSe forming a relative twist angle φ. Here, blue, magenta, and orange spheres depict Ga, In, and Se atoms, respectively. The twist angle leads to a longer range periodicity that produces a moiré potential pining excitons or photogenerated bound electron-hole pairs. (b) Moiré periodicity λ as a function of the twist angle φ for 1L-ε-GaSe/1L-γ-InSe heterostructures. (c) Intensity of the photoluminescent peak centred at ~ 1.5 eV that is associated with the interlayer exciton IX emission of a 4L-γ-InSe on 5L-ε-GaSe heterostructure as a function of the excitation energy E. Notice that the intensity of IX increases as the excitation energy surpasses 1.7 eV, the value corresponding to the intralayer exciton X₀(In) of 4L-γ-InSe but plateaus beyond ~2.1 eV the energy required to excite the X₀(Ga) of 5L-ε-GaSe. This enhancement of IX upon excitation of all exciton states confirms its interlayer nature and indicates that all**

emissions are intrinsic to the strongly coupled InSe/GaSe heterostructure[13]. (d) to (f) Photoluminescent spectra from three $n$(Ga)L-ε-GaSe/$n$(In)L-γ-InSe heterostructures having, according to second harmonic generation, twist angles $\phi = (0.3 \pm 0.5)^o$, $(23.4 \pm 1.7)^o$, and $(47.9 \pm 1.1)^o$ respectively (See, Figs. S8, S10, and S11 for the determination of $\phi$). The spectra (black markers) can be fit to a minimal set of three Gaussians (magenta, clear blue, and blue lines) describing emissions at distinct energies. Red line corresponds to their superposition. In panel d, green line depics a fit to a single Gaussian, illustrating that it does not correctly describe the data. Notice how their separation in energy $\Delta E$, given in meV and indicated by vertical dotted gray lines, increases with $\phi$. As illustrated by the sample in panel (f), regardless of the precise laser spot coordinates, every emission can be characterized by a minimum set of three energy levels, as illustrated by our fits to three Gaussians. g, Center of the Gaussians for the 5L-ε-GaSe on 4L-γ-InSe heterostructure ($\phi \cong 23.4^o$) as a function of the position, indicating that they are separated by about ~50 meV regardless of laser spot position. The scattering in energy among these points would suggest an inhomogeneous moiré lattice.

## METHODS

**Single crystal synthesis.** InSe single-crystals were grown through the Bridgman method: 6N-pure indium and 5N-pure selenium pellets in an atomic ratio of 52:48 were sealed in an evacuated quartz ampoule and subsequently placed into a radio frequency (RF) furnace where the RF power was gradually increased to raise the temperature up to 800 ºC. The ampule was then pulled through the hottest zone at a rate of 2 mm/hour. Single crystals were characterized *via* electron dispersive spectroscopy and aberration corrected transmission electron microscopy. Single crystals of GaSe were grown via a Ga flux method: Ga (6N) and Se (5N) in a ratio of 10:1 were sealed in quartz ampoule and the heated up to 1000 ºC, to be subsequently slowly cooled down to 400 ºC, and then centrifuged. The $h$-BN crystals were produced by precipitation from molten metal solutions at atmospheric pressure[64].

**X-ray diffraction.** Single crystals of GaSe were cut to appropriate sizes and mounted on glass fibers with two-part epoxy. Data were collected with a Bruker D8 Quest Kappa single crystal X-ray diffractometer with a Mo Kα IμS microfocus source, a HELIOS optics monochromator, and Photon 100 CMOS detector. Absorption corrections were performed using the Bruker SADABS program (multi-scan method)[65]. Starting crystallographic models were obtained using the intrinsic phasing method in SHELXT[66] and atomic sites were refined anisotropically using SHELXL2014[67]. X-ray diffraction confirmed that the GaSe adopts the space group 187 ($P\bar{6}m2$) with lattice parameters a = 3.7557(3) Å, and c = 15.9502(12) Å. γ-InSe adopts the $R3m$ space group (160) with cell dimensions $a$ = 4.08(2) Å and $c$ = 24.938(24) Å.

**Heterostructure fabrication.** Each constituent thin layer in the heterostructure, as well as the electrical gates, were exfoliated from bulk single-crystals onto polydimethylsiloxane(PDMS). A dry-transfer method within an Ar filled glovebox was used to assemble stacks composed of few-layer graphene/h-BN/GaSe/InSe/h-BN. The PDMS stamp containing the above heterostructure was then stamped onto a 285 nm thick $SiO_2$ layer on a $p$-doped silicon substrate. A top graphene gate was electrically contacted through Cr/Au electrodes using electron-beam lithography and electron-beam evaporation techniques.

**Photoluminescence measurements at low temperatures.** The photoluminescense measurements were performed in a Quantum Design Physical Properties Measurement System having a magnet that provides a maximum field $\mu_0 H$ = 14 T coupled to a custom probe containing a piezo stack that translates the sample under the objective of the microscope. The excitation provided by a continuous-wave laser excitation with a photon energy of 2.41 eV (514 nm wavelength) was used.The excitation light was focused through an objective (NA = 0.65) to a spot size of ~ 2 μm. The PL emission light was collected by the same objective and detected by a spectrometer equipped with a thermoelectric cooled charge-coupled device (CCD) camera. The spectra presented throughout this manuscript were collected at several temperatures and recorded with a spectral resolution of 0.11 meV.

**Time resolved photoluminescence.** Time-resolved PL measurements were performed at $T$ = 5 K using laser wavelength λ = 515 nm with a power of 500 μW. We checked that lower and higher illumination powers yielded similar lifetimes. For interlayer lifetime measurements, we excite the sample with a < 200 fs pulsed laser (Light Conversion PHAROS). Interlayer PL is spectrally filtered through a 0.5-m monochromator (Teledyne Princeton Instruments SpectraPro HRS-500), and detected with a fast time-correlated single-photon counting system composed of a fast (< 30 ps full width at half maximum) single-photon avalanche detector (Micro Photon Device PDM Series) and a picosecond event timer (PicoHarp 300).

**Second-harmonic generation (SHG) measurements.** Linearly-polarized, optical pulses with a central wavelength of 800 nm and average pulse duration of ~60 fs were generated with a Ti:sapphire laser operating at a repetition rate of 5.1 MHz. Pulses were guided through a motorized half-waveplate, and focused at normal incidence through an objective lens (20x, N.A. = 0.45) down to a $1/e^2$ spot size of ~6 μm. The incident power at the sample was measured to be ~1 mW. The sample-generated SHG light at 400 nm was collected by the same objective lens, passed through an analyzer (to select p or s output polarizations), and digitized into TTL pulses using a photomultiplier tube (Hamatsu H7360 series). Any undesired 800 nm light reflected back from the sample was removed prior to SHG detection *via* narrow-edge 400nm bandpass filters. TTL counts were recorded as a function of half-waveplate angular position, using a gated photon counter (Stanford Research Systems, SR400). All measurements were performed at room temperature and the second-harmonic nature of the detected signal was verified through the quadratic dependence of SHG intensity on the incident power.

**Twist angle determination from SHG measurements.** Twist angle values were determined from measured SHG patterns by deriving an optical response model for each material layer and fitting the resulting equations to the experimental data. More specifically, the polarization of a material layer $P_M^{(2)}$ was first related to the electric field of the incident laser $E_L$, *via*

$$P_M^{(2)} = 2\varepsilon_0 d_{i,l} \cdot E_L \cdot E_L$$

where, $\varepsilon_0$ is the permittivity of free space and $\boldsymbol{d_{i,l}}$ represents the second-order, nonlinear optical susceptibility of the material, written in contracted notation. The electric field vector of the laser $\boldsymbol{E_L}$ was further decomposed into the analyzer's basis:

$$\boldsymbol{E_L} = \begin{pmatrix} E_{L,0}\cos(\theta) \\ E_{L,0}\sin(\theta) \\ 0 \end{pmatrix}$$

Here, $E_{L,0}$ denotes the electric field strength of the laser, and $\theta$ is the angle formed between horizontal axis of the analyzer's frame of reference (i.e. p-polarized direction) and the orientation of the electric field vector. Under this assumption, the calculated SHG polar patterns exhibit four lobes, similar to the recorded signals.

In order to also account for the optical response of a second, rotated material layer, a second basis was added to the model and rotated by a twist angle $\alpha$ away from the analyzer's horizontal axis. Using the appropriate form of the $\boldsymbol{d_{i,l}}$ tensors associated with the crystallographic point groups of $\gamma$-InSe (i.e. trigonal 3m) and ε-Gase (i.e. hexagonal $P\overline{6}m2$), the polarization-dependent SHG intensities were found to be proportional to:

$$I_{InSe}^{p} \propto (d_{InSe} \cos(2\theta + \delta))^2$$

$$I_{InSe}^{s} \propto (d_{InSe} \sin(2\theta + \delta))^2$$

$$I_{GaSe}^p \propto (d_{GaSe} \cos(2\theta + \delta - 3\alpha))^2$$

$$I_{GaSe}^s \propto (d_{GaSe} \sin(2\theta + \delta - 3\alpha))^2$$

where $\delta$ was included as an additional fitting parameter to account for arbitrary crystallite orientations with respect to the laser's polarization direction.

For each sample, $\delta$ was found by moving the laser spot above a heterostructure-free region of $\gamma$-InSe, capturing a SHG polar pattern while keeping the analyzer's position fixed, and fitting the appropriate equation from our optical response model to the measured data. A similar procedure was repeated on a heterostructure-free region of $\varepsilon$-Gase, in order to determine twist angles, up to modulo 60°. The error incurred during the fitting procedure was estimated with a twist-angle-dependent, root-mean-square-error (RMSE) analysis of the form:

$$RMSE = \sqrt{\frac{\sum_{i=1}^n (y'_i - y_i)^2}{n}}$$

where $n$ represents the total number of data points, $y'_i$ are experimentally-measured SHG intensities and $y_i$ are the corresponding SHG intensity predictions given by the previous model. Each of the RMSE curves plotted against twist angles in the domain 0° to 60° exhibits a single, sharp minimum at the respective best fits for $\alpha$. The interval formed by the two twist angle values at twice the y-value of the best fit for $\alpha$ correspond to a 95% confidence interval (i.e. $2\sigma$).

Twist angles close to 0° were further differentiated from those closer to 60° by moving the laser spot to location above the heterostructure region, and scanning the SHG polar pattern for the maximum SHG response: Heterostructures with twist angles closer to 0° show an enhanced SHG response due to constructive interference, whereas those with twist angles closer to 60° destructively interfere and do not exhibit any measurable SHG signal[68]. Our measurements still exhibit an uncertainty of $2\sigma$ for twist angles close to 30°, where heterostructures display maximal SHG responses that are similar in magnitude[68].

**Theoretical band structure calculations.** Calculations were performed using the VASP code[69] using a 11x11x1 $k$-sampling grid. The energy cutoff is set to 500 eV, with a convergence criteria of $10^{-8}$ eV and for the forces to 0.01 eV/Å. In order to avoid interactions between the layers, we applied periodic boundary conditions with a vacuum space of 25 Å. We used the projector augmented wave (PAW)[70] methods with a plane wave basis. We also took into account a fractional component of the exact exchange from the Hartree-Fock (HF) theory hybridized with the density functional theory (DFT) exchange-correlation functional at the level of the range-separated HSE06 hybrid functional[71] to extract the band structures included in the manuscript.

# ASSOCIATED CONTENT

## Supporting Information

The Supporting Information is available free of charge at https://pubs.acs.org/doi/

Single-crystal x-ray diffraction pattern along the (00l) direction for a ε-GaSe single-crystal grown through a Ga flux method, from DFT calculations; Relative band alignments between 3L-ε-GaSe and 3L-γ-InSe, Electronic charge density across the γ-InSe/ε-GaSe interface, Electronic potential across the γ-InSe/ε-GaSe interface, Electric field across the γ-InSe/ε-GaSe interface, Comparison between interlayer and intra-layer exciton emissions in thick γ-InSe/ε-GaSe heterostructures, second harmonic generation and determination of the twist angle across the multiple heterostructures discussed within this manuscript, Photoluminescent emission from bulk γ-InSe, Photoluminescence spectrum from few 5L-γ-InSe at $T$ = 5 K, Photoluminescence spectrum at $T$ = 5 K for bulk ε-GaSe, Spectra of interlayer moiré emissions for several temperatures, Intensity and position in energy of the interlayer excitons as a function of the laser illumination power, possible role of the moiré potential on the intra-layer exciton emission of the ε-GaSe layer, Reproducibility of the intralayer exciton $X_0(Ga)$ emissions from ε-GaSe.

## ASSOCIATED CONTENT

**Supporting Information** - Single-crystal x-ray diffraction pattern along the (00$l$) direction for a GaSe single-crystal grown through a Ga flux method, Relative band alignments between 3L-ε-GaSe and 3L–γ-InSe, electronic charge density across the γ-InSe/ε-GaSe interface, electronic potential across the γ-InSe/ε-GaSe interface, electric field across the γ-InSe/ε-GaSe interface, comparison between interlayer and intra-layer exciton emissions in thick γ-InSe/ε-GaSe heterostructures, deconvolution of the interlayer exciton emission for the multiple heterostructures shown in Fig. 1, photoluminescence spectra from a non-annealed 4-γ-InSe/4L-ε-GaSe heterostructure, determination of the twist angle for several heterostructures through second harmonic generation, photoluminescent emission from bulk γ-InSe, photoluminescence spectrum from few 5L-γ-InSe at $T$ = 5, photoluminescence spectrum at $T$ = 5 K for bulk ε-GaSe grown via Ga flux, spectra of interlayer moiré emissions for several temperatures, intensity and position in energy of the interlayer moiré excitons as a function of the laser illumination power, observation of multiple Emissions from the intralayer emission of GaSe at the interface, possible role for the moiré potential on the intra-layer exciton emission of the ε-GaSe layer, reproducibility of the intralayer exciton $X_0(Ga)$ emissions from ε-GaSe.

## AUTHOR INFORMATION

**Corresponding author**

**Luis Balicas -** [1]*National High Magnetic Field Laboratory, Tallahassee, FL 32310, USA.* [2]*Department of Physics, Florida State University, Tallahassee, USA.* https://orcid.org/0000-0002-5209-0293; Email:lbalicas@fsu.edu

## AUTHOR CONTRIBUTIONS

W.Z. and L.B. conceived the project. W.Z. fabricated the heterostructures. W.Z., L.X., Z.L., and D.S. performed photoluminescence measurements. W.Z., F.W., and L.B. analyzed and interpreted the data. W.Z., S.M., and L.B. synthesized the γ-InSe and ε-GaSe single-crystals. S.L. and J.E. provided the h-BN crystals. J.B.F., G.T.M. and. J.Y.C. performed the structural studies. E.J.G.S. and M.A. provided the theoretical calculations. F.A.Q, A.S., and A.M.L. performed the second harmonic generation measurements and analysis to extract the twist angle. M.W. and C.H.L performed time dependent pholuminescence measurements. W.Z. and L.B. wrote the manuscript with the input of all co-authors. All authors discussed and commented on the manuscript.

## ACKNOWLEDGEMENTS

We acknowledge Justin B. Felder for assistance in collecting single crystalline x-ray data and Alex Moon for assistance in heterostructure fabrication. L.B. acknowledges support from US NSF-DMR 1807969 (synthesis, physical characterization, and heterostructure fabrication) and the Office Naval Research DURIP Grant 11997003 (stacking under inert conditions). L.B. also acknowledges the hospitality of the Aspen Center for Physics, which is supported by US NSF grant PHY-1607611. C.N.L acknowledges US NSF-DMR 1807928. J.Y.C. and J.F.B. acknowledges support from NSF DMR 2209804 and Welch Foundation AT-2056-20210327. R.B. acknowledges support from the National Science Foundation through NSF-DMR1904361. The crystal growth (S.L. and J.H.E.) in this study was supported by the Materials Engineering and Processing program of the National Science Foundation, Award No. CMMI 1538127. A. M. L. acknowledges support from the US Department of Energy (DOE), Office of Basic Energy Sciences, Division of Materials Sciences and Engineering, under contract no. DE-AC02-76SF00515. C.H.L. acknowledges support from the National Science Foundation Division of Materials Research CAREER Award No. 1945660 and the American Chemical Society Petroleum Research Fund No. 61640-ND6. EJGS acknowledges computational resources through CIRRUS Tier-2 HPC Service (ec131 Cirrus Project) at EPCC (http://www.cirrus.ac.uk) funded by the University of Edinburgh and EPSRC (EP/P020267/1); ARCHER UK National Supercomputing Service (http://www.archer.ac.uk) *via* Project d429. EJGS acknowledges the Spanish Ministry of Science's grant program "Europa-Excelencia" under grant number EUR2020-112238, the EPSRC Early Career Fellowship (EP/T021578/1), and the University of Edinburgh for funding support. This work was supported by the US-NSF (Platform for the Accelerated Realization, Analysis, and Discovery of Interface Materials (PARADIM)) under Cooperative Agreement No. DMR-2039380. The National High Magnetic Field Laboratory acknowledges support from the US-NSF Cooperative agreement Grant number DMR-1644779 and the state of Florida.

**Notes**

The authors declare no competing interests.

Supplementary Information for

# Thickness and twist angle dependent interlayer excitons in metal monochalcogenide heterostructures

*Wenkai Zheng*[§,⊥], *Li Xiang*[§,⊥], *Felipe de Quesada*[¶,£], *Mathias Augustin*[ι,ǁ], *Zhengguang Lu*[§,⊥,†], *Matthew Wilson*[ƒ], *Aditya Sood*[¶,£], *Fengcheng Wu*[∇], *Dmitry Shcherbakov*[¶], *Shahriar Memaran*[§,⊥], *Ryan E. Baumbach*[§,⊥], *Gregory T. McCandless*[ș], *Julia Y. Chan*[ș], *Song Liu*[т], *James Edgar*[т], *Chun Ning Lau*[¶], *Chun Hung Lui*[ƒ], *Elton Santos*[ι,ǁ,ơ], *Aaron Lindenberg*[¶,£], *Dmitry Smirnov*[§], *and Luis Balicas*[§,⊥]*

[§]*National High Magnetic Field Laboratory, Tallahassee, FL 32310, USA.*

[⊥]*Department of Physics, Florida State University, Tallahassee, FL 32306, USA*

[¶]*Stanford Institute for Materials and Energy Sciences, SLAC National Accelerator Laboratory, Menlo Park, CA 94025, USA*

[£]*Department of Materials Science and Engineering, Stanford University, Stanford, CA 94305, USA*

[ι]*Institute for Condensed Matter Physics and Complex Systems, School of Physics and Astronomy, The University of Edinburgh, EH9 3FD, UK.*

[ƒ]*Department of Physics and Astronomy, University of California, Riverside, CA 92521, USA.*

[∇]*School of Physics and Technology, Wuhan University, Wuhan, 430072 China.*

[¶]*Department of Physics, The Ohio State University, Columbus, OH 43210, USA.*

[ș]*Department of Chemistry and Biochemistry, Chemistry and Biochemistry Baylor University, Waco, TX 76798, USA.*

[т]*Tim Taylor Department of Chemical Engineering, Kansas State University, Manhattan, KS 66506, USA.*

[ǁ]*Higgs Centre for Theoretical Physics, The University of Edinburgh, EH9 3FD, UK.*

[ơ]*Donostia International Physics Centre, 20018, Donostia-San Sebastian, Spain.*

[†]*Present address: Department of Physics, Massachusetts Institute of Technology, Cambridge, MA 02139, USA.*

*e-mail: balicas@magnet.fsu.edu

[†]Present address: Department of Physics, Massachusetts Institute of Technology, Cambridge, MA, USA.

Figure S1. Single-crystal x-ray diffraction pattern along the (00*l*) direction for a GaSe single-crystal grown through a Ga flux method.

Figure S2. Relative band alignments between 3L-ε-GaSe and 3L–γ-InSe.

Figure S3. Electronic charge density across the γ-InSe/ε-GaSe interface.

Figure S4. Electronic potential across the γ-InSe/ε-GaSe interface.

Figure S5. Electric field across the γ-InSe/ε-GaSe interface.

Figure S6. Comparison between interlayer and intra-layer exciton emissions in thick γ-InSe/ε-GaSe heterostructures.

Figure S7. Deconvolution of the interlayer exciton emission for the multiple heterostructures shown in Fig. 1.

Figure S8. Photoluminescence spectra from a non-annealed 4-γ-InSe/4L-ε-GaSe heterostructure.

Figure S9. Determination of the twist angle between a 40L-ε-GaSe and a 60L-γ-InSe crystal through second harmonic generation.

Figure S10. Determination of the twist angle between a 4L-ε-GaSe and a 4L-γ-InSe crystal through second harmonic generation.

Figure S11. Determination of the twist angle between a 5L-ε-GaSe and a 4L-γ-InSe crystal through second harmonic generation.

Figure S12. Determination of the twist angle between a 40L-ε-GaSe and a 60L-γ-InSe crystal through second harmonic generation.

Figure S13. Determination of the twist angle between a 30L-ε-GaSe and a 50L-γ-InSe crystal through second harmonic generation.

Figure S14. Determination of the twist angle between a 15L-ε-GaSe and a 15L-γ-InSe crystal through second harmonic generation.

Figure S15. Photoluminescent emission from bulk γ-InSe.

Figure S16. Photoluminescence spectrum from few 5L-γ-InSe at $T$ = 5 K.

Figure S17. Photoluminescence spectrum at $T$ = 5 K for bulk ε-GaSe grown via Ga flux.

Figure S18. Spectra of interlayer moiré emissions for several temperatures.

Figure S19. Intensity and position in energy of the interlayer moiré excitons as a function of the laser illumination power.

Observation of multiple Emissions from interfacial Ga

Figure S20. Possible role for the moiré potential on the intra-layer exciton emission of the ε-GaSe layer.

Figure S21. Reproducibility of the intralayer exciton $X_0$(Ga) emissions from ε-GaSe comprising the 15L-γ–InSe / 15L-ε-GaSe heterostructure characterized by a twist angle of 54.6º.

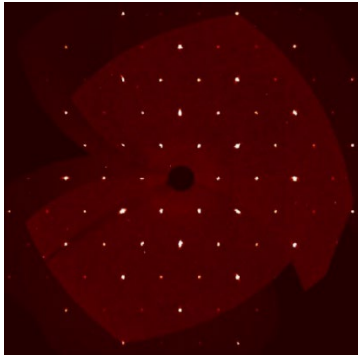

**Figure S1. Single-crystal x-ray diffraction pattern along the (00*l*) direction for a GaSe single-crystal grown through a Ga flux method.** Sharp Bragg peaks are observed indicating the absence of inter-layer mosaicity.

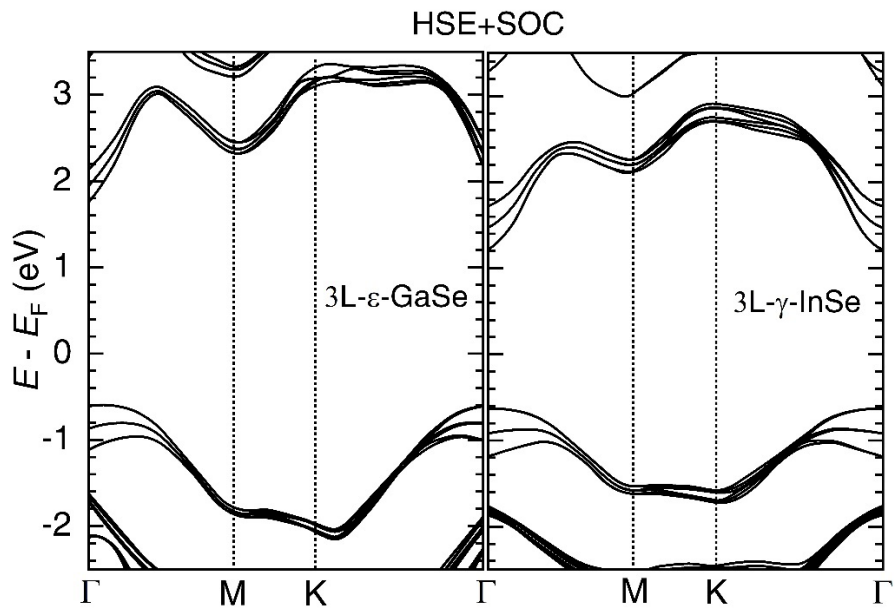

**Figure S2. Relative band alignments between 3L-ε-GaSe and 3L–γ-InSe.** Left panel: electronic band structure for 3L-ε-GaSe according to the hybrid functional HSE06 incorporating spin-orbit coupling (SOC). Right panel: electronic band structure for 3L-γ-InSe according to the hybrid functional HSE06 incorporating SOC. As the bulk and the monolayers, the tri-layers also display a type-II relative band alignment between both compounds, suggesting that this type of band alignment is maintained for most combinations in the number of layers.

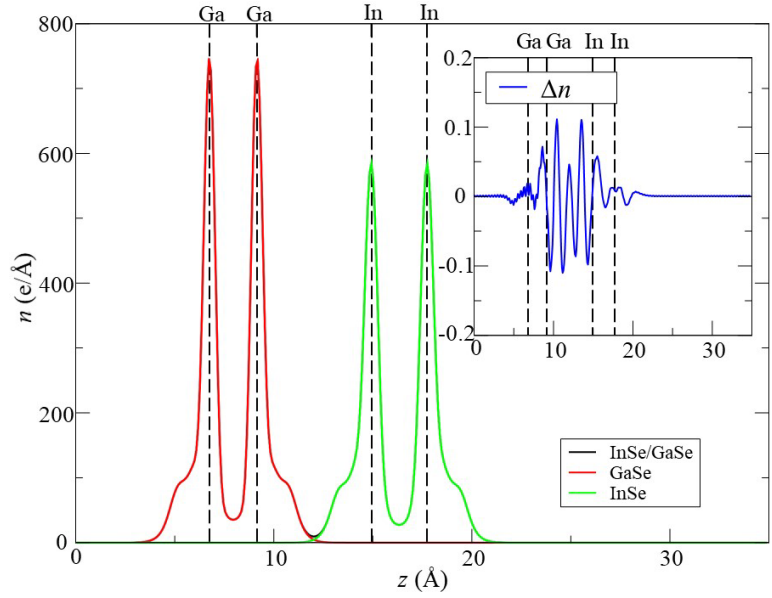

**Figure S3. Electronic charge density across the γ-InSe/ε-GaSe interface.** Electronic charge density $n$ as a function of the out-of-plane distance $z$ for the γ-InSe/ε-GaSe interface (Black), and free-standing ε-GaSe (Red) and γ-InSe (Green) monolayers. Calculations performed using vdW-corrected ab-initio simulation methods. $n$ peaks at the location of the metallic ions. Inset: Charge density difference ($\Delta n = n$(InSe/GaSe) - $n$(InSe) - $n$(GaSe), in eV/Å) as a function of $z$. Black trace is barely visible due to its overlap with the red and green traces. After the assembly of the heterostructure the largest difference in charge density $\Delta n$ becomes located between the In and Ga atoms, or in the vicinity of the Se atoms.

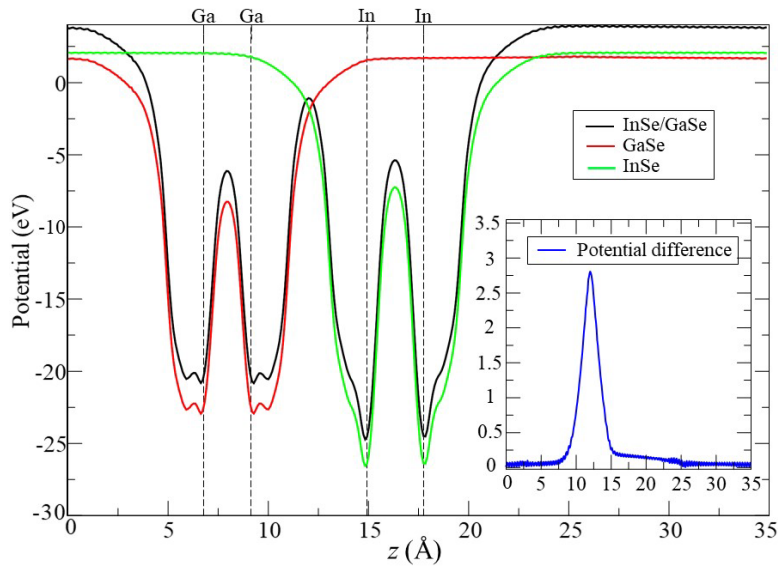

**Figure S4. Electronic potential across the γ-InSe/ε-GaSe interface.** Electronic potential Φ as a function of the out-of-plane position $z$ for the γ-InSe/ε-GaSe interface (black), and free standing ε-GaSe (Red) and γ-InSe (Green) monolayers. Calculations performed using vdW-corrected *ab-initio* simulation methods. F peaks at the location of the Ga and In atoms (vertical dashed lines). Inset: A large potential difference ($\Delta\Phi = \Phi$(InSe/GaSe) - $\Phi$(InSe) - $\Phi$(GaSe), in eV) is observed across the InSe/GaSe interface

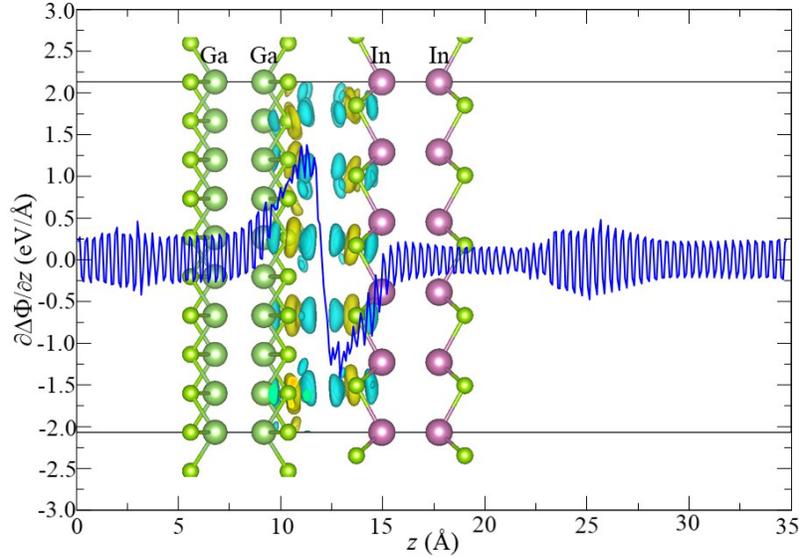

**Figure S5. Electric field across the γ-InSe/ε-GaSe interface.** Electric field *E* (blue line) across the γ-InSe/ε-GaSe interface as calculated through vdW-corrected ab-initio simulation methods, and as a function of the interlayer position *z*. Here the γ-InSe layer is assumed to be well aligned relative to the ε-GaSe one. Big purple (green) atoms depict In (Ga), and small green atoms correspond to Se.

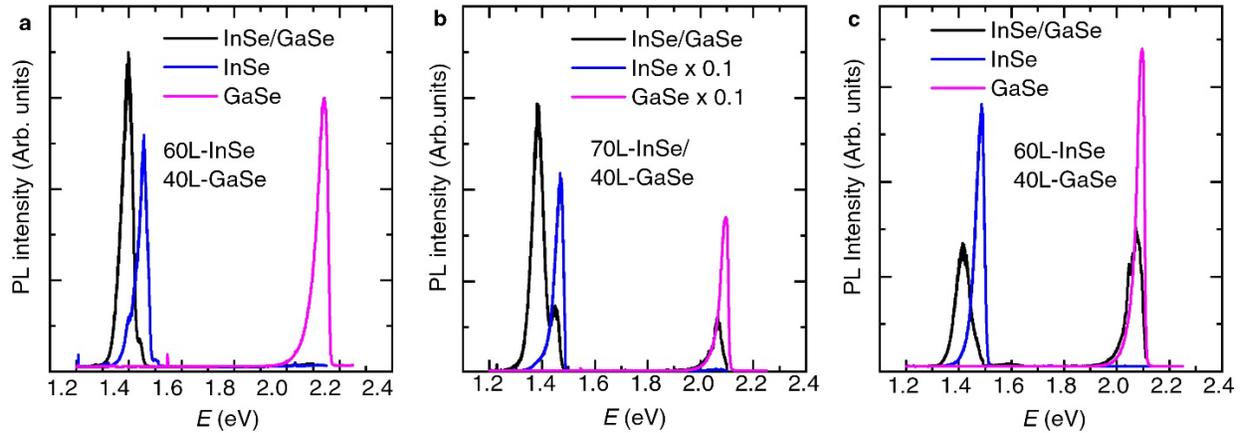

**Figure S6. Comparison between interlayer and intra-layer exciton emissions in thick γ-InSe/ε-GaSe heterostructures.** From left to right: photoluminescence spectra from three distinct heterostructures of γ-InSe/ε-GaSe composed of thick layers. Black lines correspond to spectra from the interfacial heterostructure area, blue lines to emission exclusively from γ-InSe areas, and magenta lines exclusively from the ε-GaSe areas not overlapping the γ-InSe layer. Notice the partial overlap between the interlayer exciton emission and the emission resulting from the intra-layer exciton of γ-InSe. The secondary peaks and humps seen in IX emission (black traces) at higher energies correspond to contributions from the intra-layer exciton $X_0$(In).

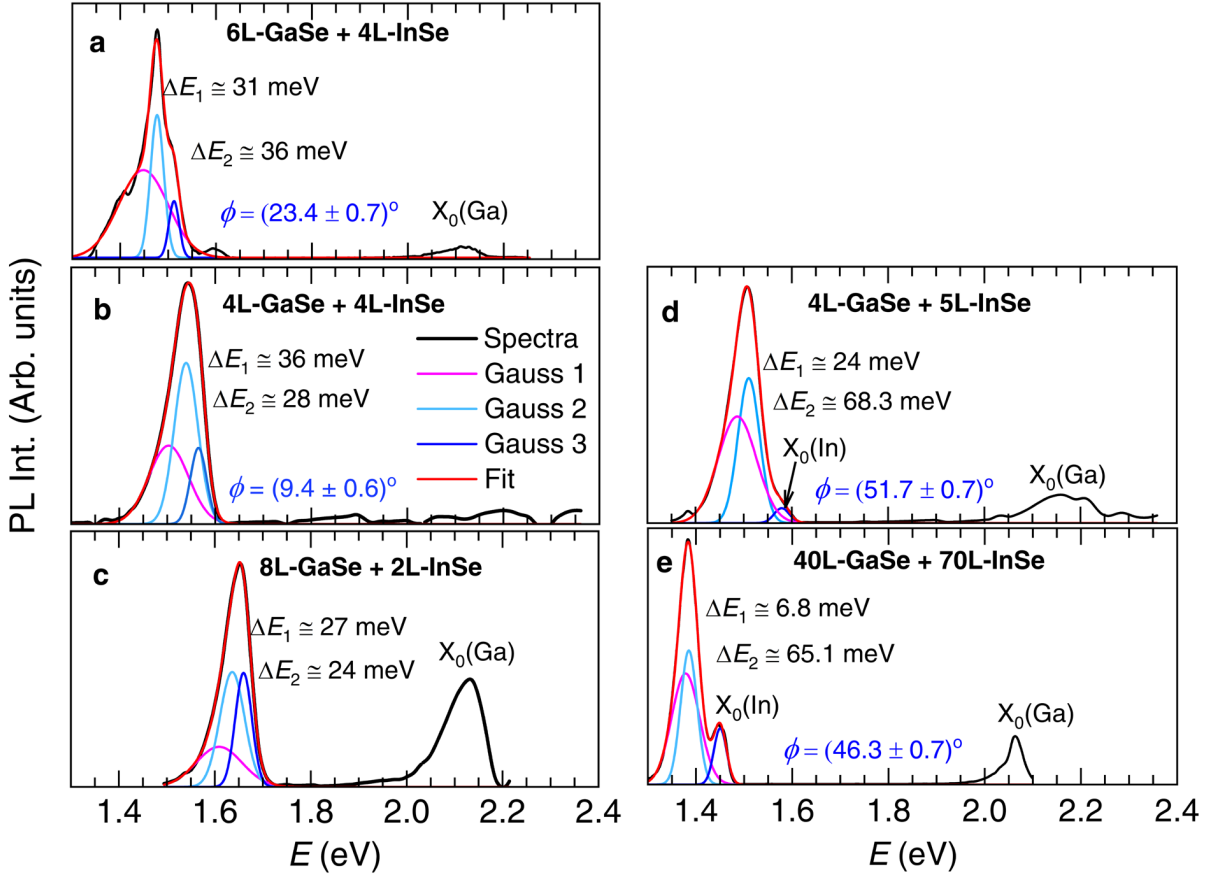

**Figure S7. Deconvolution of the interlayer exciton emission for the multiple heterostructures shown in Fig. 1.** (a) to (f) Photoluminescent spectra collected from the multiple heterostructures whose data is shown in Fig. 1(e) within the main text. Here, we fitted all IX emissions to three Gaussians, which is the minimum set to fit IX or to capture the $X_0(In)$ emission adjacent to IX in certain heterostructures. Notice, i) the clear displacement of IX towards higher energies as the number of layers decrease and ii) the larger separation in energy $\Delta E_2$ between IX and the $X_0(In)$ emission in panels (d) and (e). This non-uniformity contrasts markedly with panels (a), (b), and (c) as well as the data in Fig. 4. In panels (a), (b), (d) and (e) we include the twist angle $\phi$. As we discuss in the main text, $\Delta E$ seems to increase with the twist angle $\phi$, but a comparison between panels (d) and (e) also suggests that $\Delta E$ might be proportional to the center in energy of the IX emission. Fits for other heterostructures are shown in Fig. 4 within the main text.

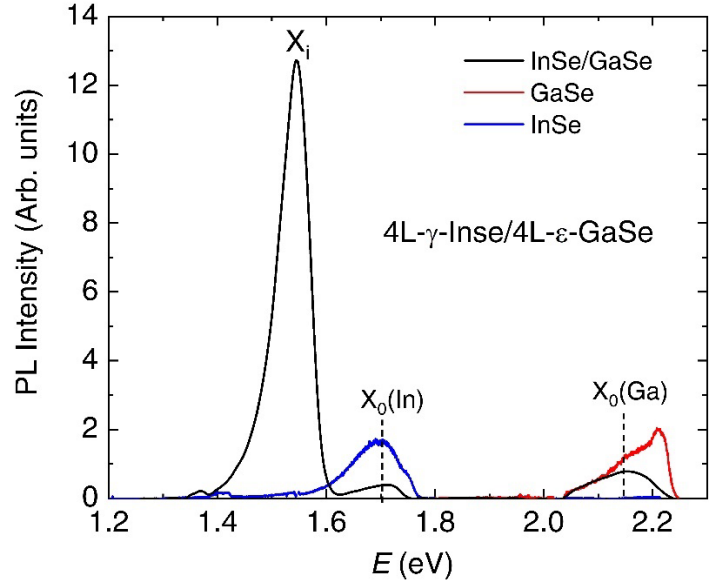

**Figure S8. Photoluminescence spectra from a non-annealed 4-γ-InSe/4L-ε-GaSe heterostructure.** Black trace corresponds to the PL spectrum from the junction area, revealing emission peaks at 1.55 eV and 1.7 eV as well as a broad peak centred at 2.15 eV. Blue trace corresponds to emission from a region of the γ-InSe layer that does not overlap the ε-GaSe one, indicating that the peak at ≈1.7 eV corresponds to the intralayer exciton $X_0$(In) of γ-InSe. Red trace is the emission from the ε-GaSe layer not in contact with γ-InSe, indicating that the peak at 2.15 eV corresponds to the intralayer $X_0$(Ga) emission. Therefore, the tall at peak at 1.55 eV ought to correspond to the interlayer IX emission.

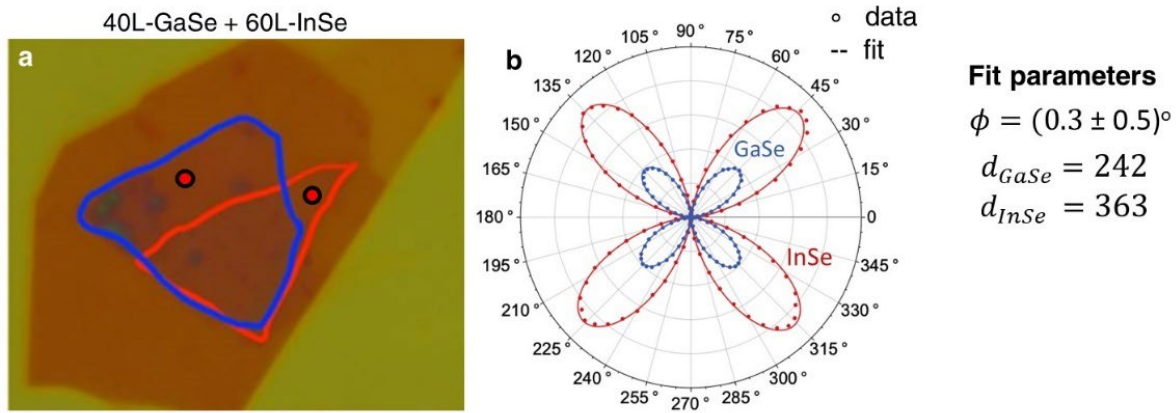

**Figure S9. Determination of the twist angle between a 40L-ε-GaSe and a 60L-γ-InSe crystal through second harmonic generation.** (a) Microphotography of the heterostructure where blue and red lines indicate the contours of the ε-GaSe and γ-InSe crystals. Red dots indicate the locations where the laser spot was focused to collect the second harmonic generation signal. (b) Amplitude of the second harmonic signal as a function of the orientation in polar coordinates. Blue and red markers correspond to experimental data while red and blue solid lines are fits to the equation discussed in the methods section which yields a twist angle $\phi = (0.3 \pm 0.5)^\circ$.

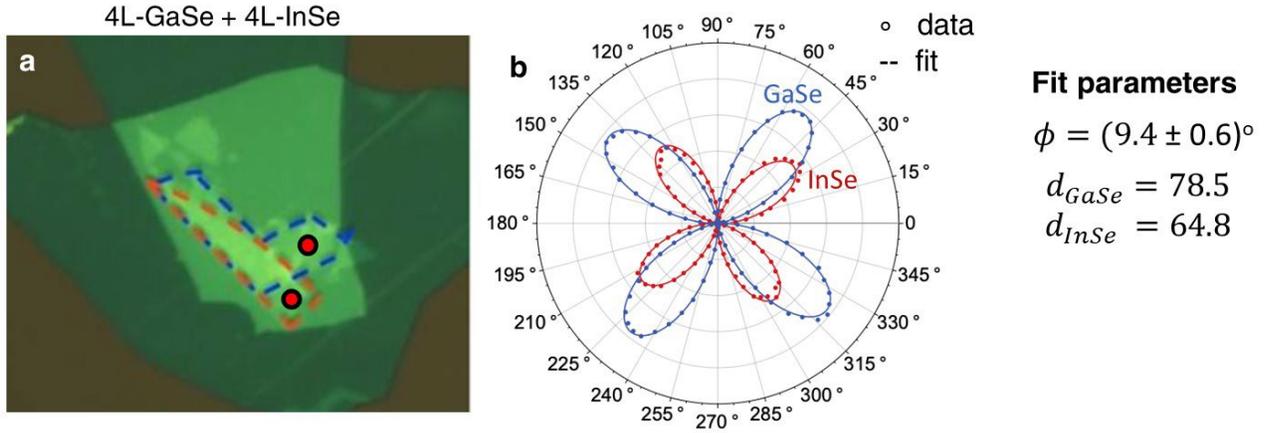

**Figure S10. Determination of the twist angle between a 4L-ε-GaSe and a 4L-γ-InSe crystal through second harmonic generation.** (a) Microphotography of the heterostructure where blue and red lines indicate the contours of the ε-GaSe and γ-InSe crystals. Red dots indicate the locations where the laser spot was focused to collect the second harmonic generation signal. (b) Amplitude of the second harmonic signal as a function of the orientation in polar coordinates. Blue and red markers correspond to experimental data while red and blue solid lines are fits yielding a twist angle $\phi = (9.4 \pm 0.6)^\circ$.

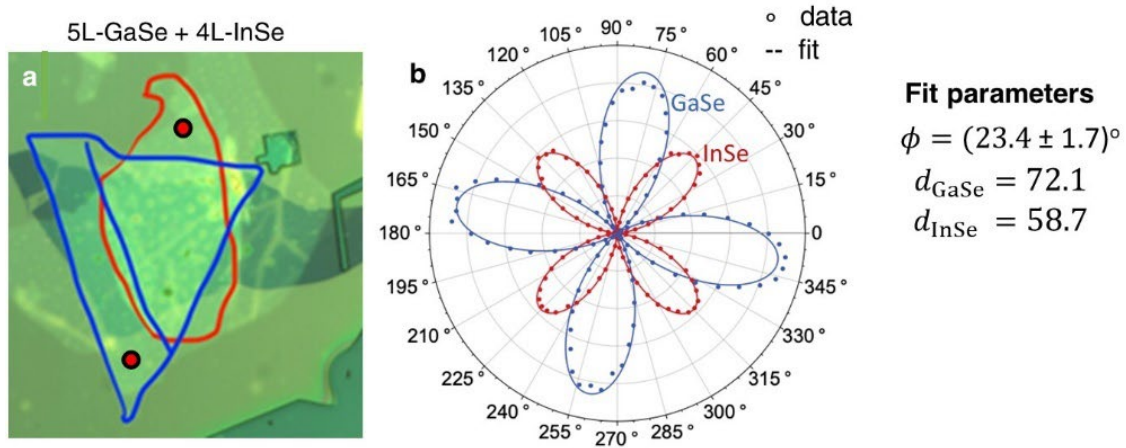

**Figure S11. Determination of the twist angle between a 5L-ε-GaSe and a 4L-γ-InSe crystal through second harmonic generation.** (a) Microphotography of the heterostructure where blue and red lines indicate the contours of the ε-GaSe and γ-InSe crystals. Red dots indicate the locations where the laser spot was focused to collect the second harmonic generation signal. (b) Amplitude of the second harmonic signal as a function of the orientation in polar coordinates. Blue and red markers correspond to experimental data while red and blue solid lines are fits yielding a twist angle $\phi = (23.4 \pm 1.7)^\circ$.

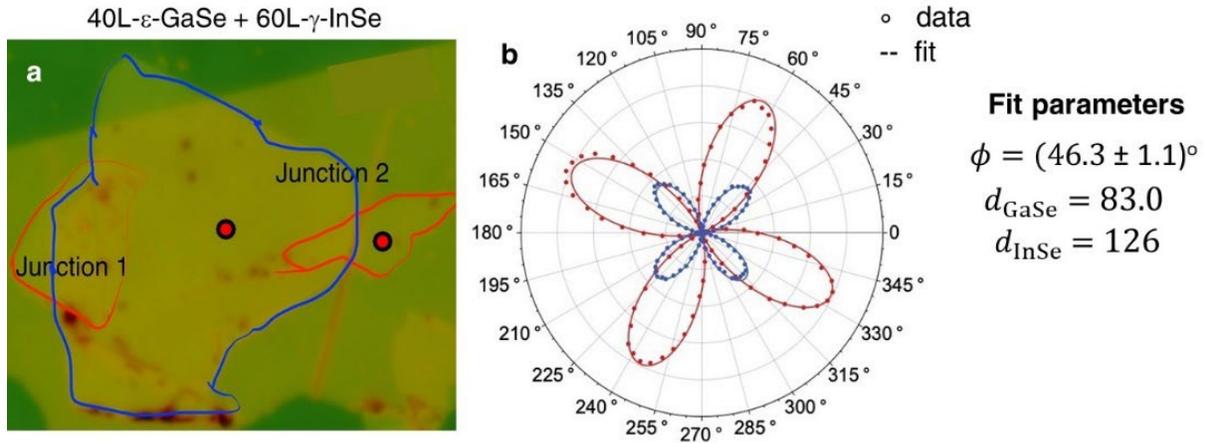

**Figure S12. Determination of the twist angle between a 40L-ε-GaSe and a 60L-γ-InSe crystal through second harmonic generation.** (a) Microphotography of the heterostructure where blue and red lines indicate the contours of the ε-GaSe and γ-InSe crystals. Red dots indicate the locations where the laser spot was focused to collect the second harmonic generation signal. (b) Amplitude of the second harmonic signal as a function of the orientation in polar coordinates. Blue and red markers correspond to experimental data while red and blue solid lines are fits yielding a twist angle $\phi = (46.3 \pm 1.1)°$.

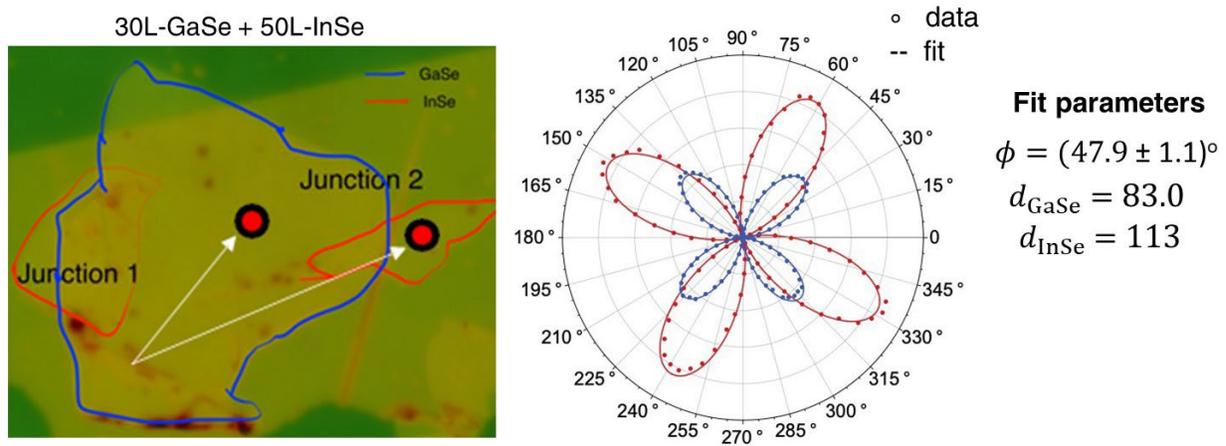

**Figure S13. Determination of the twist angle between a 30L-ε-GaSe and a 50L-γ-InSe crystal through second harmonic generation.** (a) Microphotography of the heterostructure where blue and red lines indicate the contours of the ε-GaSe and γ-InSe crystals. Red dots indicate the locations where the laser spot was focused to collect the second harmonic generation signal. (b) Amplitude of the second harmonic signal as a function of the orientation in polar coordinates. Blue and red markers correspond to experimental data while red and blue solid lines are fits yielding a twist angle $\phi = (47.9 \pm 1.1)°$.

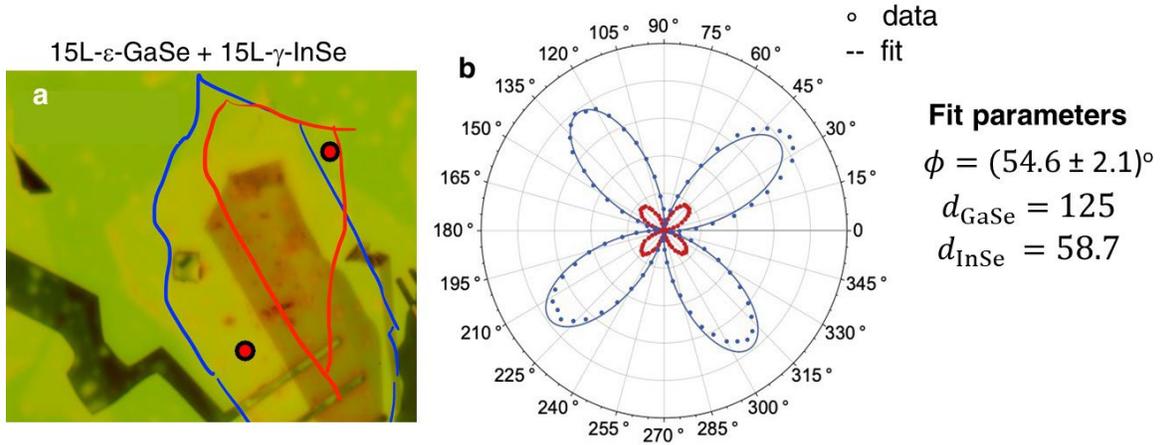

**Figure S14. Determination of the twist angle between a 15L-ε-GaSe and a 15L-γ-InSe crystal through second harmonic generation.** (a) Microphotography of the heterostructure where blue and red lines indicate the contours of the ε-GaSe and γ-InSe crystals. Red dots indicate the locations where the laser spot was focused to collect the second harmonic generation signal. (b) Amplitude of the second harmonic signal as a function of the orientation in polar coordinates. Blue and red markers correspond to experimental data while red and blue solid lines are fits yielding a twist angle $\phi = (54.6 \pm 2.1)°$.

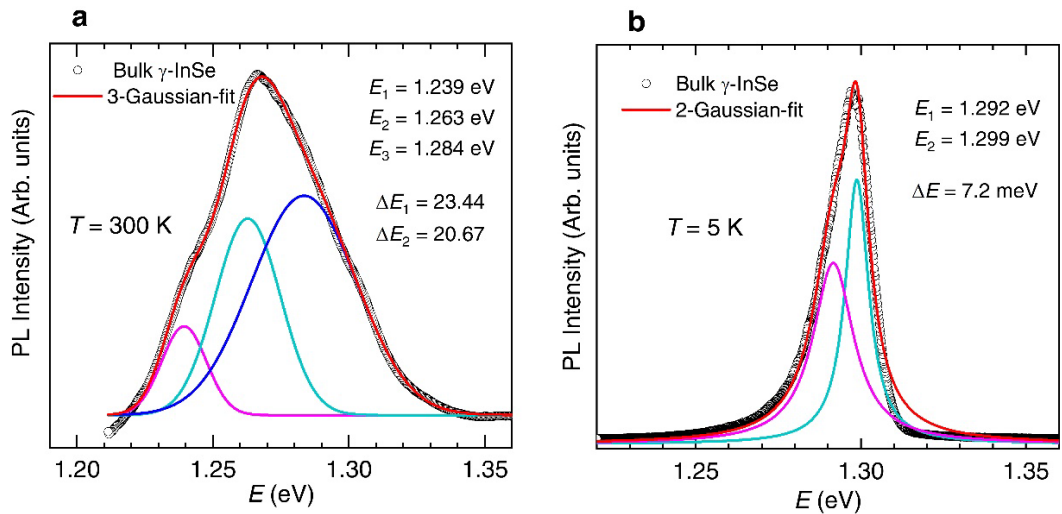

**Figure S15. Photoluminescent emission from bulk γ-InSe.** (a) As reported by other groups (Ref. [46] in the main text) the PL emission of our bulk InSe single-crystal at $T = 300$ K, grown via the Bridgman technique, displays multiple emissions due, for example, to phonon replicas centred at 1.239 eV, 1.263 eV, and 1.284 eV, respectively. Notice i) that their position in energy $E$ as well as their separation in energy $\Delta E$ do not coincide with any of the emissions in Figs. 2d, 2e, and 2f in the main text, and ii) that the amplitude of these emissions scale with $E$, in contrast to what is observed in the γ-InSe/ε-GaSe heterostructures. (b) Photoluminescent emission from bulk γ-InSe collected at $T = 5$ K. Notice that i) the centre of the emission is displaced to higher energies, and ii) the PL peak narrows considerably as the amplitude of the phonons decrease. Although, the peak is relatively well-described by a single Lorentzian, it can be improved with the addition of a second one. However, the resulting energy separation is just $\sim 7$ meV. Therefore, we can firmly discard that the PL emissions collected from γ-InSe / ε-GaSe heterostructures result from bulk γ-InSe.

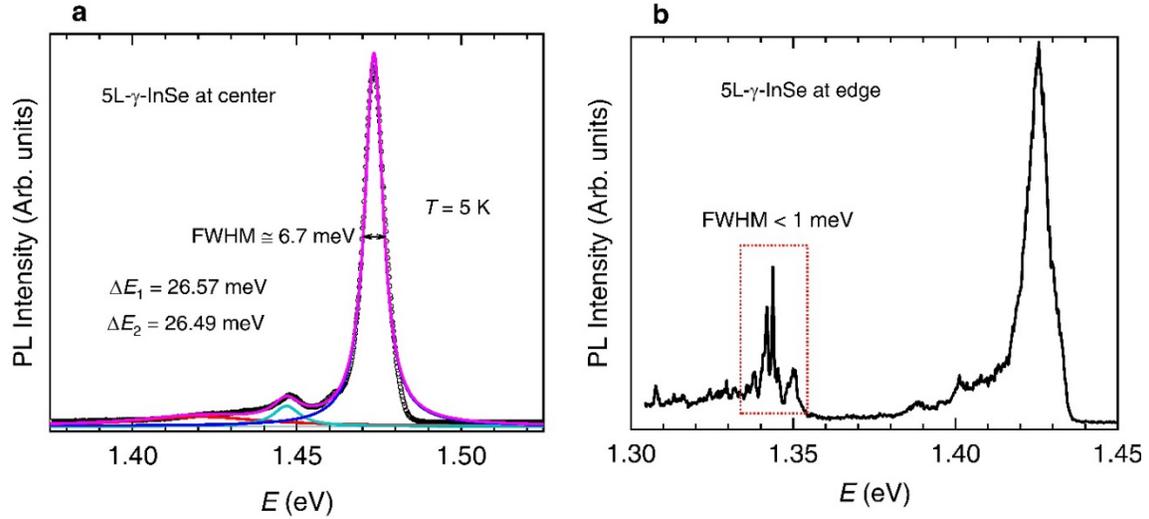

**Figure S16. Photoluminescence spectrum from few 5L-γ-InSe at $T$ = 5 K.** (a) PL spectrum (markers) from the centre of a 5L-γ-InSe crystal encapsulated among *h*-BN layers in a double graphene gate configuration. Magenta line corresponds to the superposition of three Lorentzians (blue, green, and red lines) to fit the spectrum. The resulting fitted peaks are equally spaced in energy with an energy separation $\Delta E \sim 26.5$ meV with this value being rather distinct from the D$E$ values shown in Fig. 2 within the main text for the γ-InSe / ε-GaSe heterostructures. Notice also that the main peak full width at half maximum is only 6.7 meV in contrast to what is shown in Fig 2 of the main text. (b) PL spectrum collected near the boundary or lateral edge of the crystal, where an accumulation of defects resulting from the exfoliation process is observed. Notice the emerge of a set of very narrow peaks (enclosed by red square) around and below 1.35 eV associated with defect emission. Their width and location in energy contrasts markedly with the interlayer exciton emissions IX described within the main text.

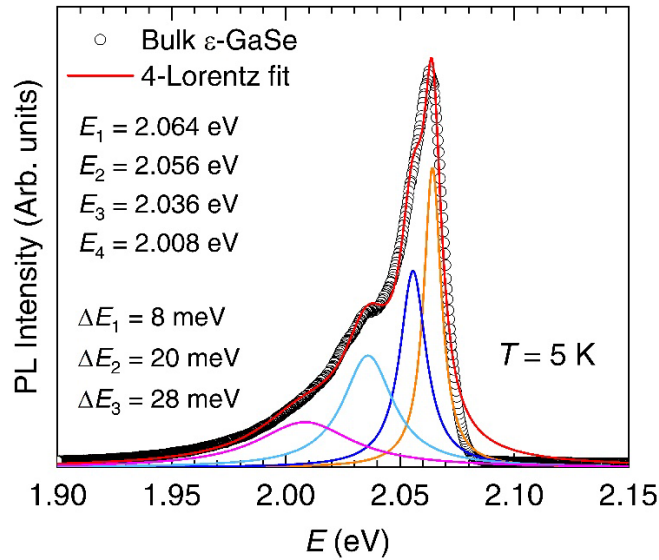

**Figure S17. Photoluminescence spectrum at $T$ = 5 K for bulk ε-GaSe grown via Ga flux.** Markers depict the PL spectrum**,** magenta, clear blue, blue and orange lines are fits to Lorentz functions, while the red line corresponds to their superposition. In marked contrast with the spectra shown in Fig. 4 within the main text, the amplitude of the individual Lorentz functions and their separation $\Delta E$ increases as a function $E$.

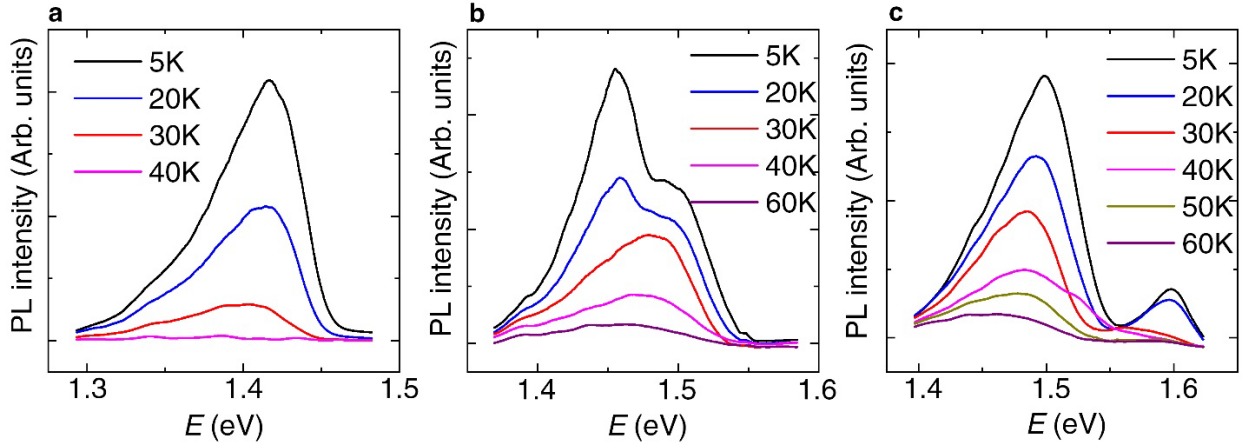

**Figure S18. Spectra of interlayer moiré emissions for several temperatures.** (a), (b), and (c), IX emissions centered at 1.4, 1.45 and 1.5 eV, respectively. The increase in temperature suppresses all IX emissions suggesting a prominent role for phonon scattering.

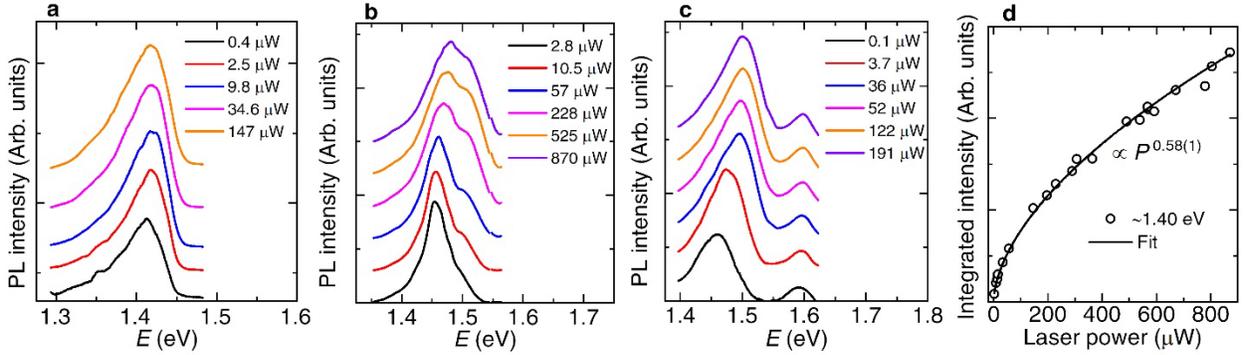

**Figure S19. Intensity and position in energy of the interlayer moiré excitons as a function of the laser illumination power.** (a) Renormalized PL spectra associated with the IX emission at ~1.4 eV as a function of the laser illumination power. (b) Renormalized PL spectra associated with the IX emission centered at ~1.45 eV (at a distinct spatial location in the heterostructure) as a function of the laser illumination power. (c) Renormalized PL associated with the IX emission centered at ~1.5 eV (at yet another spatial location in the heterostructure) as a function of the laser illumination power. Notice that IX at both $E \cong 1.45$ eV and $E \cong 1.5$ eV slightly blue shifts as the power increases while the emission at $E \cong 1.6$ eV does not point to interactions among densely populated excitons. (d) Integrated intensity $I$ for the main peak at 1.4 as a function of laser power, indicating that they it follows a power law $I \propto P^{\alpha}$ with $\alpha < 1$. All emissions display similar values implying a tendency to saturate at higher laser powers. And this is compatible with an interlayer exciton emission that tends to saturate at higher laser power due to Coulomb like dipole-dipole interactions

## Observation of multiple Emissions from the intralayer emission of GaSe at the interface

The intrinsic intralayer exciton emission $X_0(Ga)$ of 5L-ε-GaSe, that is describable by a single Lorentzian (Fig. S19a), displays at the interface behavior akin to IX: it becomes describable by several superimposed emissions that are nearly equispaced in energy (Fig. S19b, c). Fitting the emission from a heterostructure characterized by $\phi \cong 23.4^\circ$ to simple Lorentzians yields an average energy separation $\Delta E \cong 22$ meV. But as the twist angle $\phi$ increases to $46.3^\circ$ or to $54.6^\circ$ (Fig. S19c, and Figs. S10, S11 and S13 for the determination of $\phi$), $\Delta E$ decreases to average values of ~14 and 11 meV, respectively. This decrease in $\Delta E$ is consistent with the expected increase in $\lambda$ as $\phi$ approaches 60°. This implies that $X_0(Ga)$ is remarkably susceptible to the moiré potential even for samples as thick as 40L-ε-GaSe on 60L-γ-InSe, or that the $X_0(Ga)$ exciton would seem to be strongly elongated along the $z$-direction[50]. Perhaps due to an enhancement in the interlayer charge transfer[13] we could not observe $X_0(Ga)$ for $\phi < 23.4°$. $\Delta E$ ~22 meV  (Fig. S17b) is much smaller than $\Delta E \sim 50$ meV (Fig. 2e) measured for IX from this same sample. This is expected given that intralayer and interlayer excitons are subjected to distinct moiré potential profiles[11]. We emphasize that the amplitude of these $X_0(Ga)$ emissions as function of $E$, or their separation $\Delta E$, do not match those of bulk ε-GaSe at low $T$s (Fig. S16) and are surprisingly reproducible (albeit not perfectly) among disting locations throughout the heterostructure (Fig. S20). This cannot be reconciled with defects, impurities or imperfections, whose emissions would be random or strongly location dependent. Therefore, we are led to conclude that our observations are akin to those in Ref.[3] which attributes the splitting of the A exciton emission of $WSe_2$ (into three peaks) to the role of the moiré potential. It is however, remarkable that this effect is still detected in heterostructures built from thick layers, but this will require further studies.

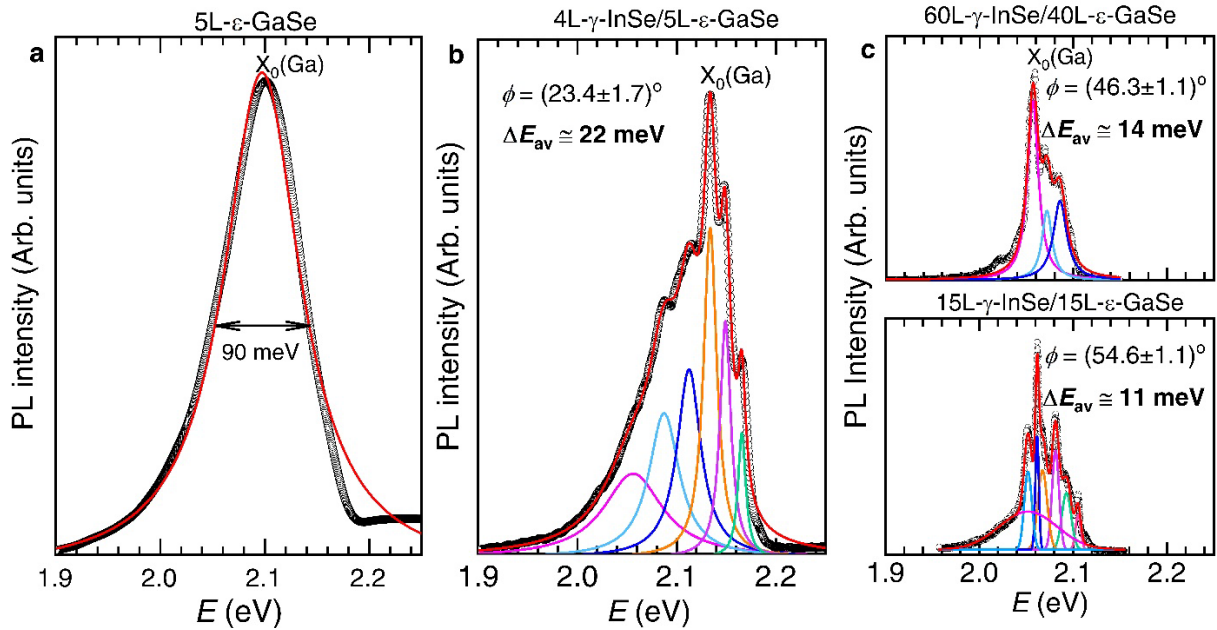

**Figure S20. Possible role for the moiré potential on the intra-layer exciton emission of the ε-GaSe layer.** (a) PL emission as a function of the energy $E$ from the Intra-layer exciton $X_0(Ga)$ of a 5L-ε-GaSe collected at $T = 5$ K from an area outside the interface with γ-InSe. In contrast to bulk ε-GaSe, the PL emission can be reasonably well-fitted to a single Lorentz function.  (b), $X_0^{Ga}$ PL emission from the same 5L-ε-GaSe crystal at the interface with 4L-γ-InSe with both layers forming a twist angle $\phi = (23.4 \pm 1.7)°$ according to second harmonic generation (Supplementary Fig. 10). In the interfacial region, the original single PL peak from 5L-ε-GaSe becames describable by multiple emissions, as indicated by fits to several Lorentzian distributions (coloured lines). Red line corresponds to their superposition. Their separation D$E$ decreases as $E$ increases displaying an average value $\Delta E_{av} \cong 22$ meV. (c), Top panel: multiple emissions are observed in all interfaced crystals including a 40L-ε-GaSe crystal interfaced with 60L-γ-InSe forming a twist angle $\phi = (46.3 \pm 1.1)°$. In this case, the PL spectra can be fitted to three Lorentzians (coloured lines, with the red line corresponding to their superposition). For this larger value of $\phi$, we obtain a lower value $\Delta E_{av} \cong 14$ meV. Lower panel: IX emission for a 15L-ε-GaSe on 15L-γ-InSe forming a twist angle $\phi = (54.6 \pm 21)°$ (Supplementary

Fig. 12) whose emissions display an average energy separation $\Delta E_{av} \cong 11$ meV. Hence, $\Delta E_{av}$ for $X_0$(Ga) progressively decreases as $\phi$ approaches 60º. This correlation between $\phi$ and $\Delta E_{av}$ indicates that the moiré potential can also alter the $X_0$(Ga) emission even in thick GaSe layers.

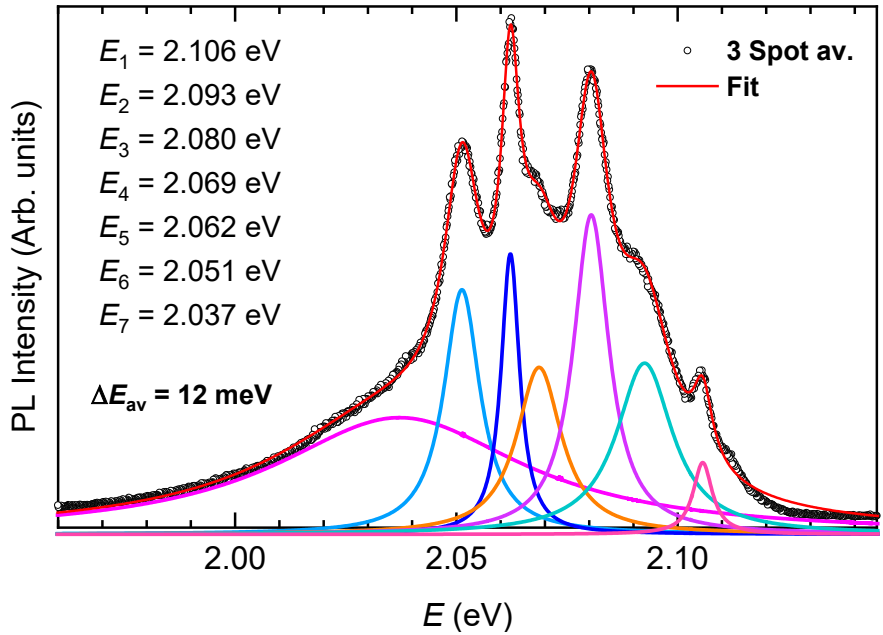

**Figure S21. Reproducibility of the intralayer exciton $X_0$(Ga) emissions from ε-GaSe comprising the 15L-γ−InSe / 15L-ε-GaSe heterostructure characterized by a twist angle of 54.6º.** Markers depict the average PL spectrum collected from three well-separated, and distinct locations along the heterostructure whose PL data (from another location) is displayed in the lower panel of Fig. 4c of the main text. Coloured lines are Lorentz functions fits to multiple emissions, while the red line represents their superposition. Although the overall envelope of the PL emission is location dependent, the centre of the individual peaks (indicated by $E_n$) or emissions is reproducible among the distinct locations. This cannot be reconciled with defects, impurities or imperfections, whose emission would be strongly location dependent. Notice also that this reproducibility contrasts markedly with the lack of reproducibility reported in Ref. [3].